\documentclass[10pt,letterpaper]{article}
\usepackage[top=0.85in,left=2.75in,footskip=0.75in]{geometry}

\usepackage{amsmath,amssymb}

\usepackage{changepage}

\usepackage[utf8x]{inputenc}

\usepackage{textcomp,marvosym}

\usepackage{cite}

\usepackage{nameref,hyperref}

\usepackage[right]{lineno}

\usepackage{microtype}
\DisableLigatures[f]{encoding = *, family = * }

\usepackage[table,dvipsnames]{xcolor} 

\usepackage{array}

\newcolumntype{+}{!{\vrule width 2pt}}

\newlength\savedwidth

\newcommand\thickhline{\noalign{\global\savedwidth\arrayrulewidth\global\arrayrulewidth 2pt}%
\hline
\noalign{\global\arrayrulewidth\savedwidth}}

\raggedright
\usepackage[aboveskip=1pt,labelfont=bf,labelsep=period,justification=raggedright,singlelinecheck=off]{caption}

\makeatletter
\renewcommand{\@biblabel}[1]{\quad#1.}
\makeatother

\usepackage{lastpage,fancyhdr,graphicx}
\usepackage{epstopdf}
\fancyheadoffset[L]{2.25in}
\fancyfootoffset[L]{2.25in}
\usepackage{lscape}
\usepackage{multirow}
\usepackage{rotating}

\begin{document}
\vspace*{0.2in}

\begin{flushleft}
{\Large
\textbf\newline{Inference using a composite-likelihood approximation for stochastic metapopulation model of disease spread}
}
\newline
\\
Gaël Beaunée\textsuperscript{1*},
Pauline Ezanno\textsuperscript{1},
Alain Joly\textsuperscript{2},
Pierre Nicolas\textsuperscript{3\Yinyang},
Elisabeta Vergu\textsuperscript{3\Yinyang}
\\
\bigskip
\textbf{1} Oniris, INRAE, BIOEPAR, 44300 Nantes, France
\\
\textbf{2} GDS Bretagne, France
\\
\textbf{3} Université Paris-Saclay, INRAE, MaIAGE, Jouy-en-Josas, France
\\
\bigskip

%
%
\Yinyang These authors contributed equally to this work.





* gael.beaunee@inrae.fr

\end{flushleft}
\section*{Abstract}

Spatio-temporal pathogen spread is most often partially observed at the metapopulation scale. Available data correspond to proxies and are incomplete, censored and heterogeneous. Moreover, representing such biological systems often leads to complex stochastic models. Such complexity together with data characteristics make the analysis of these systems a challenge.

Our objective was to develop a new inference procedure to estimate key parameters of stochastic metapopulation models of animal disease spread from longitudinal and spatial datasets, while accurately accounting for characteristics of census data. We applied our procedure to provide new knowledge on the regional spread of \emph{Mycobacterium avium} subsp. \emph{paratuberculosis} (\emph{Map}), which causes bovine paratuberculosis, a worldwide endemic disease.

\emph{Map} spread between herds through trade movements was modelled with a stochastic mechanistic model. We used comprehensive data from 2005 to 2013 on cattle movements in 12,857 dairy herds in Brittany (western France) and partial data on animal infection status in 2,278 herds sampled over the period 2007 - 2013. Inference was performed using a new criterion based on a Monte-Carlo approximation of a composite likelihood, coupled to a numerical optimization algorithm (Nelder-Mead Simplex-like).

Our criterion showed a clear superiority to alternative ones in identifying the right parameter values, as assessed by an empirical identifiability on simulated data. Point estimates and profile likelihoods highlighted that a very large proportion ($>0.80$) of the dairy cattle herds were infected in 2005 with a low infection prevalence on average. We found a moderate and stable risk of purchasing infected cattle from outside the metapopulation ($0.14$). We confirmed the low average sensitivity of the diagnostic test ($0.21$).

Our inference procedure could easily be applied to other spatio-temporal infection dynamics, especially for long-lasting endemic diseases. It is of particular interest when ABC-like inference methods fail due to difficulties in defining relevant summary statistics.

\section*{Author summary}

\textcolor{blue}{
}


\section*{Introduction}

    To better understand endemic pathogen spread and identify appropriate and targeted control measures, it is necessary to account for interactions occurring between the farm and the regional scales. Indeed, farms are not isolated populations. Especially, movements of animals connect close or distant farms, and represent one of the major transmission routes at large scale~\cite{Bajardi_2012caa,Buhnerkempe_2014fv}. Then, as a result of the well-known rescue effect, pathogen spread occurring due to between-herd contacts may induce infection persistence at large scale while fadeout occurs locally~\cite{ezanno2020epidemics}. Furthermore, the diversity of cattle farming systems and local specificities of herd management (e.g. herd size, renewal rate, contact structure) also influence pathogen spread~\cite{Kunzler2014,Lindstrom2011}. Each herd has its own demography, which influences the risk of acquiring an infectious disease, the severity and duration of outbreaks within the herd, and the risk of transmitting pathogens to other herds. Factors such as herd size~\cite{BrooksPollock_2009kg}, renewal practices~\cite{Stahl_2008cj}, within-herd contact structure~\cite{Ezanno_2008fz, Marce_2011ic}, biosecurity practices~\cite{Flaten_2005hd}, and geographical location~\cite{Ersboll_2010jp} have been highlighted as largely influencing pathogen spread. Hence, a detailed representation of both the within-population demographic processes (with heterogeneity in herd management among herds) and the epidemiological processes is needed to ensure simulations being realistic enough for tackling practical issues.

    However, this gives rise to an increase in model complexity, and thus also in the number of parameters involved and which have to be accurately specified to ensure robust predictions. In managed populations (e.g. in livestock), high resolution data (such as the European databases recording comprehensively cattle movements~\cite{Dutta_2014hx}) are available. This allows a fine grain calibration of demographic processes and to adequately represent population dynamics~\cite{Beaunee:2015ed,qi2019}. Conversely, there is a paucity of available epidemic data, and epidemic dynamics are mostly observed as punctual data points. In addition to being incomplete (data are available for only part of the population), epidemic data generally correspond to proxies, processes being only partially observed. Key events of the infection processes are not observed. For example, if the status of an animal can be known at specific time points (e.g. test dates), the moment when it has changed is not observed. In addition, the difficulty of accurately calibrating a complex model is reinforced by the uncertainty of observations as associated with the imperfect sensitivity of diagnostic tests and more generally infection detection issues.

    There is a need for innovative methods to reconcile complex mechanistic models with sparse and heterogeneous data. On the one hand, both the characteristics of such complex models (which are dynamic, with a large number of variables, and even stochastic) and of available data (which are spatiotemporal, incomplete, censored, and imperfect) prevent from defining the likelihood and, therefore, make it impossible to use classical estimation methods (i.e. maximum likelihood). On the other hand, recent advances in likelihood free methods, such as Approximate Bayesian Computations (ABC)~\cite{beaumont2002approximate,Toni:2009jt}, and the increase in performance and availability of computing resources have enabled fitting complex models to large-scale epidemic data~\cite{CHISSTER200921,BrooksPollock:2014if}. However, building the appropriate summary statistics (i.e. containing enough information to be a discriminating criterion), and thus to use ABC-like methods, remains a challenge when tackling heterogeneous units (such as holdings), each having data of variable accuracy and at variable time steps. In addition, although these methods have the advantage of being simulation-based, they generate of a huge number of particles, even in the case of ABC-SMC (Sequential Monte Carlo)~\cite{Toni:2009jt} which is one of the most effective.

    Bovine paratuberculosis, also known as Johne's disease, is a relevant example of an enzootic disease distributed worldwide~\cite{MarcelABehr_2010uv}, inducing large economic losses and management issues to farmers~\cite{Anonymous_YCCDV071,Lombard2005vt,Garcia_2015bx}, and for which heterogeneous and sparse epidemiological data are available at farm scale thank to surveys implemented by animal health services in some areas. There is a need for more robust predictions of the spatio-temporal spread of the causative agent of this chronic inflammatory bowel disease, \emph{Mycobacterium avium} subsp. \emph{paratuberculosis} (\emph{Map}), in order to assess and compare the range of possible control options currently available and to identify the most relevant ones. Many different models have been proposed to predict Map spread at farm scale~\cite{Marce_2010kt}, only one focusing on the regional scale~\cite{Beaunee:2015ed} to tackle the issue of implementing targeted and adapted control strategies~\cite{beaunee2017controlling}. Because of the characteristics of the pathogen, the associated disease, and the host population, and despite an effort to render such models as parsimonious as possible, these models are quite complex and suffer from parameter uncertainty, especially as regards Map transmission, diagnostic test sensitivity, and knowledge about current prevalence at both animal and farm scales.

    The objective of this work is twofold. The first objective is to reconcile complex mechanistic epidemiological models and sparse and heterogeneous observed data by developing an alternative approach to the ABC methods to provide an appropriate method when summary statistics are difficult to define, and allowing to better account for unsampled herds in the infection dynamics and to more accurately account for the characteristics of census data for animals tested as positive. The second objective is to gain new knowledge on Map spread at a regional scale by estimating the key parameters of a metapopulation epidemiological model from longitudinal and spatial data collected in Brittany (Western France). To the best of our knowledge, this is the first time such an issue is tackled using this type of data. We propose here a new use of these data, taking into account their specific characteristics (scattered, uncomplete, ...) to infer unobservable parameters.

    To achieve our objectives, (i) we updated an existing epidemiological model of the regional Map spread, (ii) we defined an relevant criterion and an appropriate estimation procedure, (iii) we assessed through simulations the accuracy (closeness to the true value) and precision (variability) of the proposed method, and (iv) we applied our method on real serological survey data to infer seven parameters of the regional epidemiological model of Map spread among cattle farms.

\section*{Materials and methods}


\subsection*{Epidemiological data \label{subsubsectionEpidemiologicalData}}

Among the 18 french administrative regions, Brittany has the highest
dairy production and accounted for 22.6\% of the total milk volume
produced in 2018 in France (source: DRAAF Bretagne). Animal health
services (GDS Bretagne) collect data on paratuberculosis in Brittany
since the early 2000s using individual serological tests (ELISA) of
very high specificity but low and, in fact, relatively unknown
sensitivity. Most of these tests take place in punctual screenings
that encompass, when possible, all animals that are at least two-year
old in the herd. These herd-level screenings can follow the detection
of an infected animal (typically after observing clinical signs of the
disease or after the positive result of a diagnostic test prescribed
for trade), or they can be carried out in the context of a
certification program implying regular tests. Although sampling was
thus not totally random, we considered here the data as containing
unbiased information on herd-level prevalence at heterogeneous points
in space and time.

\begin{figure}[!t]
\begin{adjustwidth}{-2.25in}{0in} 
\centering 
\includegraphics[width=0.85\paperwidth]{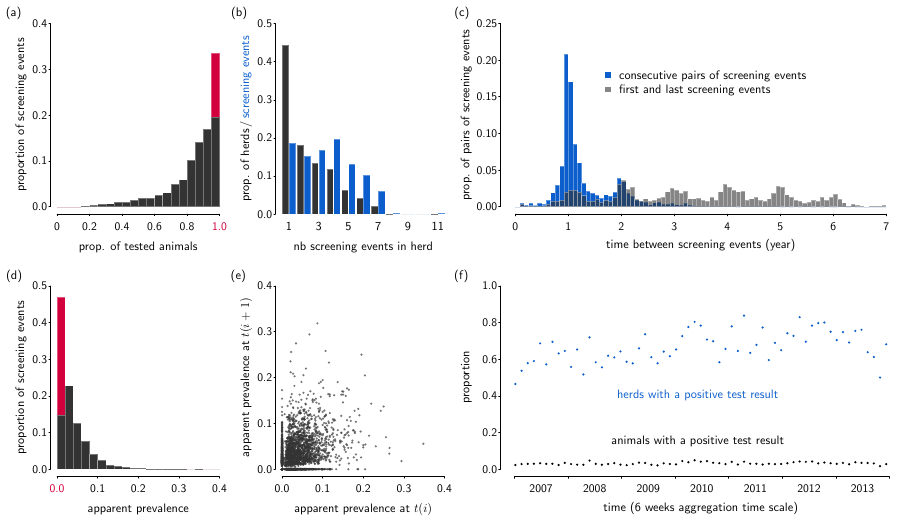}
\caption{{\bf Overview of the serological data (5,434 selected screening events)
  available for monitored herds in Brittany (western France) between 2007 and 2013.}
  Panels of top row describe the testing scheme: (a) proportion of animals of more than two years
  in the herd tested per screening event; (b) number of screening
  events in herd for each herd (black) or for each screening event (blue);
  (c) time between pairs of screening events in a same herd (consecutive
  screening events in blue, first and last screening events in gray),
  seasonal pattern, due to higher number screening events in
  winter. Panels of bottom row describe the results of the
  serological tests: (d) apparent prevalence among all screening
  events; (e) comparison of apparent prevalence of two consecutive screening events;
  (f) apparent prevalence over time: proportion of test-positive animals in tested herds ({\sl i.e.} average prevalence, in
  black) and of herds with at least one test-positive animal (blue).}
\label{fig:dataepidemio}
\end{adjustwidth} 
\end{figure}

GDS records, which contained the result of each individual test
along with precise identification of the animal, were processed to
build a high quality dataset of measures of prevalence aggregated at
herd level over a 7-year period spanning from 2007 to 2013. For
this purpose, tests made on females at least two-year old were grouped
into "screening events'' consisting of all the tests made in a window
of four weeks. The date of the screening event was set as the day with
the largest number of tests. Some screening events were discarded,
either because herds belonged to holdings not included in the
metapopulation model built on available animal movement data (see
below) or because they covered less than 30 animals and 50\% of the
herd (a minimum of 10 animals was imposed in all cases). This lead to
a dataset of 5,434 screening events distributed in 2,278 herds with
an average proportion of animals in the herd tested per screening
event of 85.3\% (Fig~\ref{fig:dataepidemio}a).
Most of the screening events (81.4\%) belonged to series of 2
or more for a same herd (Fig~\ref{fig:dataepidemio}b). Typical time
between consecutive screening events in the same herd was 1 year, with
up to 7 years between the first and the last screening event
(Fig~\ref{fig:dataepidemio}c).

Results of the serological tests established an average apparent
prevalence as low as 3.2\%, but with at least one positive test in as
much as 67\% of the screening events
(Fig~\ref{fig:dataepidemio}d). Apparent prevalence measured at
consecutive screening events displayed a mild, but statistically
significant, level of correlation (Pearson's r=0.39, p-value
$<2.2e-16$, Fig~\ref{fig:dataepidemio}e). Plots of apparent prevalence
over time did not reveal any clear trend of temporal evolution
(Fig~\ref{fig:dataepidemio}f). We refer below to these serological
results as the observed epidemiological trajectory. From it, we
propose to draw inference on the true (unobserved) prevalence and on
within-herd transmission based on comparison with the stochastic
outcomes of a simulation model.

\subsection*{Simulation model}

The regional spread of Map was simulated using a stochastic
metapopulation model adapted from
\cite{Beaunee:2015ed,beaunee2017controlling} that accounts for
within-farm and between-farm dynamics. The main characteristics of
this model which encompassed here 12,857 farms and the modifications made for this study are
described below.

\subsubsection*{Within-herd model}

\begin{figure}[!t]
  \centering 
  \includegraphics[width=0.99\textwidth]{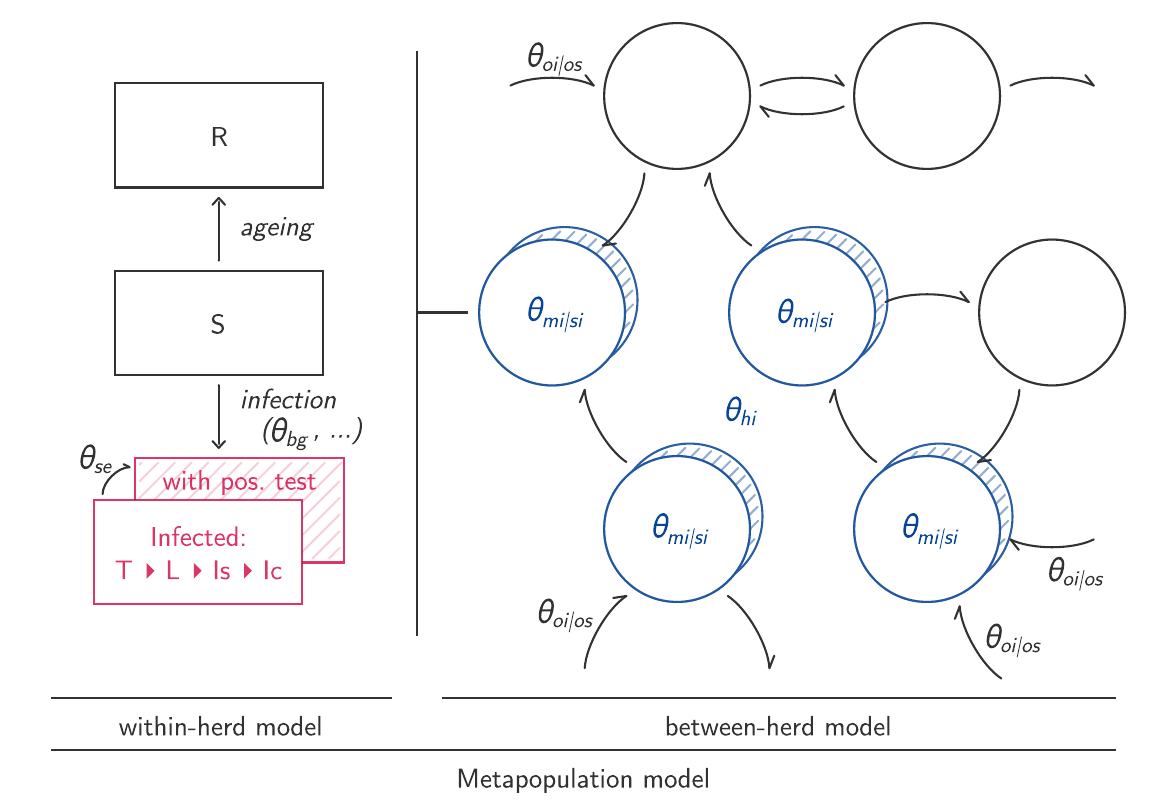} 
\caption{{\bf Conceptual model of Map spread within and between
    herds.}  The left panels gives a simplified view of the
  within-herd model; $S$, $R$, $T$, $L$, $Is$, and $Ic$ are the six
  main compartments of the within-herd model. The right panel
  corresponds to the between-herd model. Curved arrows represent
  animal movements between herds, herds in blue are infected at the
  onset of the simulation. In both panels, compartments with hatched
  line correspond to animals with a positive test
  result. Definitions of the parameters are given in Table~\ref{tab:paramdef}.}
  \label{figMetapopulationModelParatub}
\end{figure}

Within-herd transmission of Map is represented by a stochastic
compartmental model simulated with a discrete time step of one
week. The compartments, illustrated in the left panel of
Fig~\ref{figMetapopulationModelParatub}, distinguish six mutually
exclusive classes of animals: young animals susceptible to infection
(\emph{S}); older animals (after one year) no longer
susceptible (\emph{R}); transient shedders after infection (\emph{T});
latently infected but not infectious (\emph{L}); moderately
infectious (\emph{Is}); highly infectious and clinically affected (\emph{Ic}).

Transmission is modelled as occurring through five routes: (i)
vertically \emph{in utero}, (ii and iii) horizontally through the
ingestion of contaminated colostrum or milk, and (iv and v) indirectly
via a contaminated local environment (contaminated by shedding calves)
or with the general environment of the farm (contaminated by all
shedders). Animals in states $T$, $Is$ and $Ic$ shed \emph{Map} in
their faeces, and thus contaminate the farm environment. $L$
corresponds to a state between $T$ and $Is$ in which animals are
barely detected as shedders and their shedding is thus neglected~\cite{Nielsen_2006ft,Mitchell_2015eo}.
Adult infection (after one year of age) is
possible~\cite{Anonymous_KdLMcx46,Windsor_2010cb} but is very rare and
is also neglected.

Mortality and culling are modelled as stochastic processes whose rates
are calibrated specifically for each herd. Compared to
\cite{Beaunee:2015ed,beaunee2017controlling}, the model incorporates
also a new type of compartment to account for the premature culling of
animals detected as positive by serological tests. Namely, each of the
existing compartments is split into two sub-compartments that
distinguish animals tested positive from untested animals or
animals tested negative.

Screening events are simulated at points in space (herds) and time
matching the observed epidemiological trajectory. For each screening
event, a number of animals given by the proportion of tested animals
are randomly selected (among animals of two years and more in the
considered herd). The result of each individual serological test is
then drawn as the outcome of a Bernoulli trial with probability of
success $\theta_{se}$ if the selected animal is infected ({\it i.e.} in
compartment $T$, $L$, $Is$ or $Ic$) and is negative for a non-infected
animal, assuming a perfect specificity of the test. After a positive test,
the average time before culling was set to 24 weeks (as calibrated on
real data).

\subsubsection*{Between-herd model}

Animal movements between herds are implemented deterministically to
reproduce the date, origin, destination, and age of actually exchanged
animals (see data from cattle tracing records below). The compartment
of each exchanged animal is randomly selected according to the
distribution of the population in the source herd (excepted for
animals in compartment $Ic$ which are not allowed to move).

This previously published regional model was modified to allow fixing,
using dedicated parameters, the probability of purchasing an infected
animal from outside the metapopulation. When such an infected animal
is purchased, its precise compartment ($T$, $L$, or $Is$) is drawn
randomly according to the distribution of infected animals in the
whole metapopulation.

\subsubsection*{Animal movements data \label{section_mvt_data}}

Animal movements were extracted from the French cattle identification
database. This database records the life history of each animal from
birth to death, including its stays in different types of holdings
(i.e. farms, markets, and assembling centers). The cattle trade
network was simplified down to farm-to-farm movements, neglecting
short sojourns of animals in other types of holdings which are not
expected to cause new \emph{Map} infections.

Only movements of females of dairy and crossed breeds were considered
since French dairy cattle is composed almost exclusively of females,
with breeding based on artificial insemination.  Fattening activities
involving other breeds that are most often conducted in a different
building or area of the farm were neglected, and thus we consider each
farm as a single herd. Herds with less than 15 dairy females in total
or 5 dairy females over two years old, were not included in the network,
as these farms are unlikely to represent dairy production units.

Over the 9-year period considered in the simulation (from 2005 to
2013), the resulting metapopulation encompassed 12,857
farms, involved 919,304 animal
movements, among which 24.4\% were between farms belonging to the
metapopulation, 28.8\% from external holdings and 46.8\% to external
holdings.

\begin{figure}[!t]
\begin{adjustwidth}{-2.25in}{0in} 
\centering 
\includegraphics[width=0.85\paperwidth]{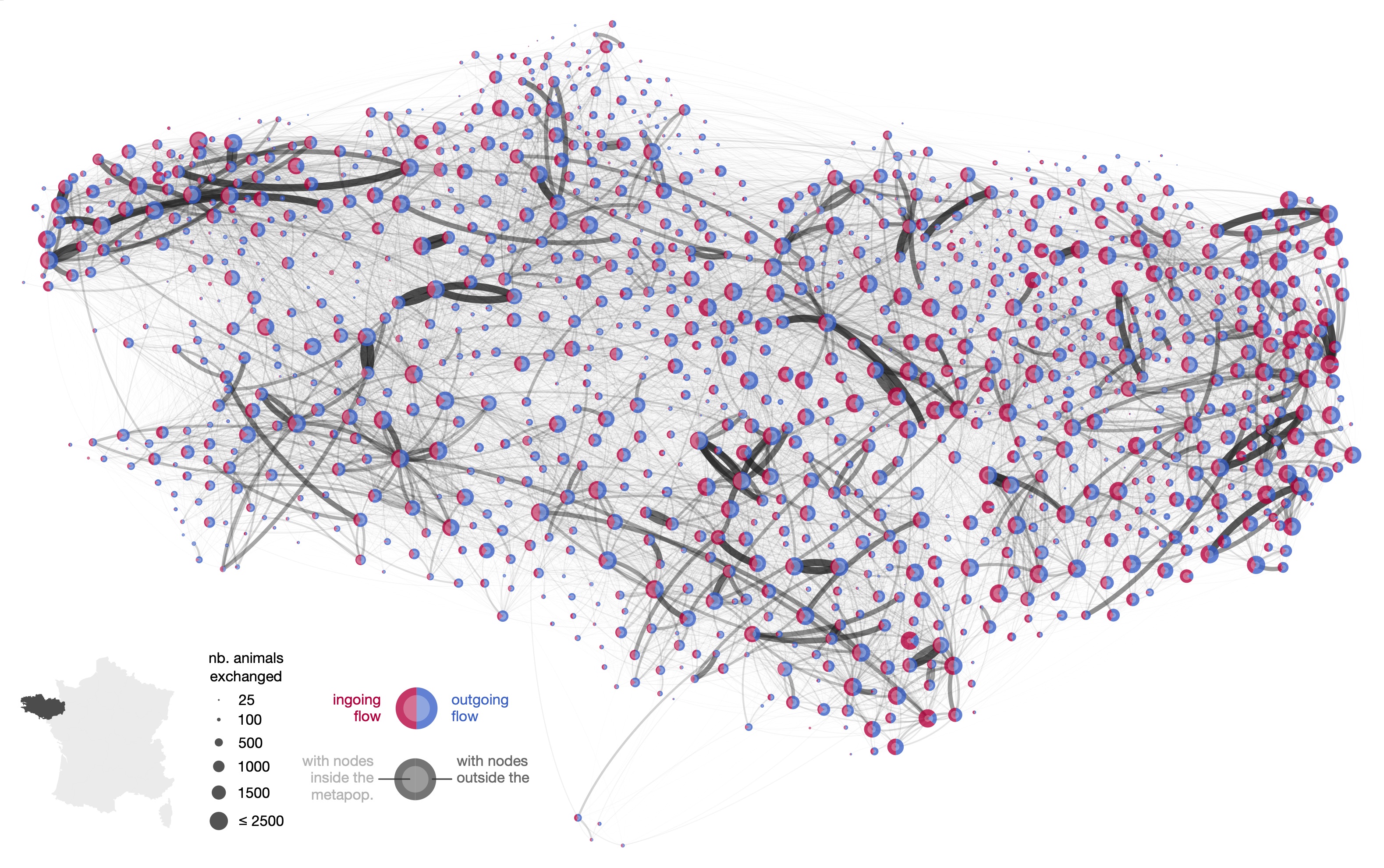}
\caption{{\bf Cattle trade network between dairy herds located in
    Brittany (western France) for the period 2005-2013.} Data has been
  aggregated at the commune level for this representation. Each
  commune is represented on the map by a disc whose size reflects
  the total number of exchanged animals. Exchanges between communes are
  represented by curved lines with thickness proportional to the
  number of animals exchanged over the period. Each disc is divided in two parts
  according to the ingoing flow (red) and the outgoing flow (blue), which are
  themselves split in an inner part corresponding to exchanges with nodes inside
  the metapopulation, and an outer part corresponding to exchanges with nodes
  outside the metapopulation.}
\label{fig:brittanynetwork}
\end{adjustwidth} 
\end{figure}

\subsection*{Inference method}

\subsubsection*{Parameter space}

 Table~\ref{tab:paramdef} gives the meaning and allowed range of
 parameter values for the seven key parameters of the epidemiological
 model (Fig.~\ref{figMetapopulationModelParatub}) which are
 estimated. The set of these seven parameters is denoted
 $\theta=(\theta_{\mathit{hi}},\theta_{\mathit{mi}},\theta_{\mathit{si}},\theta_{\mathit{oi}},\theta_{\mathit{od}},\theta_{\mathit{bg}},\theta_{\mathit{se}})$. Of
 note, parameters $\theta_{\mathit{mi}}$ and $\theta_{\mathit{si}}$
 are those of a normal distribution whose support is truncated to
 $(0.0,0.9)$ to describe the actual initial within-herd prevalence in
 infected herds. When helpful for interpretation, we report the
 parameters of the corresponding truncated normal distribution,
 denoted $\theta_{\mathit{mi}^\mathit{t}}$ and
 $\theta_{\mathit{si}^\mathit{t}}$.


\begin{table}[t]
\begin{adjustwidth}{-2.25in}{0in} 
        \setlength\extrarowheight{2pt}
\centering
\caption{
{\bf Definition of the seven parameters of the epidemiological model to be estimated.}}
\begin{tabular}{llrr}
\thickhline
Model component and parameter definition & Notation\textsuperscript{a} &  Min\textsuperscript{b} & Max\textsuperscript{c} \\
\thickhline
Initial distribution of infected herds in the metapopulation in 2005 & & & \\
 \hspace{1cm}Proportion of infected herds & $\theta_{\mathit{hi}}$ & $0.0$ & $1.0$ \\ 
 \hspace{1cm}Distribution of within-herd prevalence in infected herds & & 0.0 & 0.9 \\
 \hspace{2cm} Mean of non-truncated Gaussian distribution & $\theta_{\mathit{mi}}$ &  $-1.0$ & $1.0$ \\
 \hspace{2cm} Standard deviation of non-truncated Gaussian distribution &$\theta_{\mathit{si}}$ & $0.0$ & $1.0$ \\
 \hspace{2cm} Mean and standard deviation of the corresponding truncated distribution &$\theta_{\mathit{mi}^t}$,$\theta_{\mathit{si}^t}$ &  &  \\
\hline
Probability of purchasing infected cattle from outside the metapopulation & & & \\
 \hspace{1cm} Initial probability & $\theta_{\mathit{oi}}$ & $0.0$ & $1.0$ \\
 \hspace{1cm} Change over the whole period & $\theta_{\mathit{od}}$ &  $-1.0$ &  $1.0$ \\
\hline
Within-herd transmission through the general environment of the farm & & &\\
 \hspace{1cm} Coefficient govering transmission rate (log10 scale)  & $\theta_{\mathit{bg}}$ & $-8.0$ & $-4.0$ \\
\hline
Diagnostic test & & &\\
\hspace{1cm} Sensitivity & $\theta_{\mathit{se}}$ &  $0.0$ & $1.0$ \\
\thickhline
\end{tabular}
\begin{flushleft} \textsuperscript{a} $\theta_{\mathit{mi}^t}$ are $\theta_{\mathit{si}^t}$ are deduced from $\theta_{\mathit{mi}}$ are $\theta_{\mathit{si}}$ and thus do not belong to the list of seven free parameters. \textsuperscript{b} and \textsuperscript{c} gives endpoints of open intervals for the seven estimated parameters and of the allowed range for the within-herd prevalence in infected herds ({\it i.e.} support of truncation for the Gaussian distribution).
\end{flushleft}
\label{tab:paramdef}
\end{adjustwidth}
\end{table}

\subsubsection*{Approximate composite likelihood}

The whole observed epidemiological trajectory is denoted $y$. It
consists of the number of positive and negative tests, respectively
denoted $y_{i,j,+}$ and $y_{i,j,-}$
($y_{i,j,\cdot}=y_{i,j,+}+y_{i,j,-}$), for each herd $i$ and screening
event $j$. Given $\theta$, the simulation
algorithm allows to generate random trajectories consisting not only
of the positive/negative outcomes of the tests, denoted $\tilde{y}$, but
also of variables that were not directly observed such as the exact
number of infected/non infected animals, denoted $\tilde{x}$.

Inference proceeds by optimization over the parameter space of an
objective function that compare a given observed trajectory $y$ to a
set of $K$ trajectories simulated with parameters $\theta$. Evaluation
of the objective function is necessarily noisy since it depends on a
simulated sample of $K$ random trajectories, denoted
$(\tilde{x}^{(k)},\tilde{y}^{(k)})_{k=1\ldots K}$. The function that
was finally retained is built as the logarithm of a
\underline{c}onditional \underline{c}omposite likelihood (CC), denoted
$\ell_{CC}(\theta; y)$, whose terms are approximated using the $K$
simulated trajectories. The approximation will be referred to as
$\tilde{\ell}^K_{CC}(\theta; y)$ for its noisy evaluation, and
$\ell^K_{CC}(\theta; y)$ for its expected value. The conditional composite
likelihood accounts for a first-order Markov dependency between
observed values from a same herd. It writes as
\begin{eqnarray}
  \ell_{CC}(\theta; y) & = & \sum_{i} \biggl[ \log \pi
    (y_{i,1}|\theta) + \sum_{j > 1} \log \pi
    (y_{i,j}|y_{i,j-1},\theta)\biggl]\,,  \label{eq:conditionalcompositelikelihood}
\end{eqnarray}
and its approximation writes
\begin{eqnarray}
  \tilde{\ell}^K_{CC}(\theta; y) & = & \sum_{i} \biggl[ \log \tilde{\pi}_K
    (y_{i,1}|\theta) + \sum_{j > 1} \log \tilde{\pi}_K
    (y_{i,j}|y_{i,j-1},\theta)\biggl]\,,  \label{eq:approxconditionalcompositelikelihood}
\end{eqnarray}
where $\pi(\cdot|\theta)$ denotes densities under the exact model and
$\tilde{\pi}_K(\cdot|\theta)$ denotes their approximations.

The objective function $\ell_{CC}(\theta; y)$ is a composite
likelihood in that it is written as a likelihood but it does not
account for all the dependencies of the true stochastic
model. Further neglecting dependencies in the formulation of the
composite likelihood, we also considered optimizing a
\underline{m}arginal \underline{c}omposite likelihood (MC),
\begin{eqnarray}
  \ell_{MC}(\theta; y) & = & \sum_{i,j} \log \pi
    (y_{i,i}|\theta) \,,  \label{eq:marginalcompositelikelihood}
\end{eqnarray}
whose approximation writes
\begin{eqnarray}
  \tilde{\ell}^K_{MC}(\theta; y) & = & \sum_{i,j} \log \tilde{\pi}_K (y_{i,j}|\theta) \,.  \label{eq:approxcompositelikelihood}
\end{eqnarray}

The $\tilde{\pi}_K(y_{i,j}|\theta)$'s are obtained by Monte-Carlo
integration as the marginal density of $y_{i,j}$ under a simplified
binomial model for the observed test results $y_{i,j}$ given the
unobserved proportion of infected animals $x_{i,j,+}/x_{i,j,\cdot}$
and the sensitivity of the diagnostic test $\theta_{se}$. Denoting
$\tilde{\psi}_{i,j}^{(k)}$ the corresponding expected proportion of
positive tests,
\begin{eqnarray}
  \tilde{\psi}_{i,j}^{(k)} & = & \theta_{se}\frac{\tilde{x}_{i,j-1,+}^{(k)}}{\tilde{x}_{i,j-1,\cdot}^{(k)}}\,, \label{eq:psi}
\end{eqnarray}
we write
\begin{eqnarray}
  \tilde{\pi}_K (y_{i,j}|\theta) & \propto & \frac{1}{K} \sum \limits_{k=1}^{K} {\left(\tilde{\psi}_{i,j}^{(k)}\right)}^{y_{i,j,+}} {\left(1-\tilde{\psi}_{i,j}^{(k)}\right)}^{y_{i,j,-}}\,, \label{eq:approxcompositelikelihood_marginalterm}
\end{eqnarray}
where the proportionality relationships accounts for a binomial term
not needed when comparing composite likelihoods since it depends only
on $y_{i,j}$ and contributes as a same additive constant into
$\tilde{\ell}^K_{CC}(\theta; y)$ and $\tilde{\ell}^K_{MC}(\theta; y)$.
The same approach is used on the joint density $(y_{i,j-1},y_{i,j})$
to obtain the first-order Markov terms in $\ell^K_{CC}(\theta; y)$ as
\begin{eqnarray}
  \tilde{\pi}_K (y_{i,j}|y_{i,j-1},\theta) & = & \frac{\tilde{\pi}_K (y_{i,j-1},y_{i,j}|\theta)}{\tilde{\pi}_K (y_{i,j-1}|\theta)}\,,
\end{eqnarray}
with
\begin{eqnarray}
  \lefteqn{\tilde{\pi}_K (y_{i,j-1},y_{i,j}|\theta)}\\
  & \propto & \frac{1}{K} \sum\limits_{k=1}^{K} {\left(\tilde{\psi}_{i,j-1}^{(k)}\right)}^{y_{i,j-1,+}} {\left(1-\tilde{\psi}_{i,j,-1}^{(k)}\right)}^{y_{i,j-1,-}} {\left(\tilde{\psi}_{i,j}^{(k)}\right)}^{y_{i,j,+}} {\left(1-\tilde{\psi}_{i,j}^{(k)}\right)}^{y_{i,j,-}}\,. \label{eq:approxcompositelikelihood_conditionalterm} \nonumber
\end{eqnarray}

Eq.~\ref{eq:approxcompositelikelihood_marginalterm} would give
$\tilde{\pi}_K(y_{i,j}|\theta)=0$ when $y_{i,j,+}>0$ but
$\tilde{x}^{(k)}_{i,j,+}=0$ for all $k=1\ldots K$. To obtain a
strictly positive value also in these rare cases, the value of
$\tilde{\pi}_K(y_{i,j}|\theta)$ was taken as
        \begin{eqnarray}
        \tilde{\epsilon}_K(y_{i,j}|\theta) = \frac{1}{K} {\left(\frac{\theta_{\mathit{se}}}{\bar{\tilde{x}}_{i,j,\cdot} }\right)}^{y_{i,j,+}} {\left(1-\frac{\theta_{\mathit{se}}}{\bar{\tilde{x}}_{i,j,\cdot}} \right)}^{y_{i,j,-}}\,,
        \end{eqnarray}
where $\bar{\tilde{x}}_{i,j,\cdot}$ is the average of
$\tilde{x}^{(k)}_{i,j,\cdot}$ across the $K$ simulated
trajectories. The rationale for the choice of this
$\tilde{\epsilon}_K(y_{i,j}|\theta)$ is to correspond to one infected
animal at point $(i,j)$ in one of the $K$ simulated
trajectory. Similarly, $\tilde{\pi}_K(y_{i,j}|y_{i,j-1},\theta)$ could
not be computed with
Eq.~\ref{eq:approxcompositelikelihood_conditionalterm} when
$y_{i,j-1,+}>0$ but $\tilde{x}^{(k)}_{i,j-1,+}=0$ for all $k=1\ldots K$
and this term was in these rare cases replaced by
$\tilde{\pi}_K(y_{i,j}|\theta)$ computed with
Eq.~\ref{eq:approxcompositelikelihood_marginalterm}.

\subsubsection*{Alternative objective functions}

Other objective functions were considered in the initial
steps of this work but not used for final inference. Three are more
particularly mentioned in the results section. The first is the
\underline{a}verage \underline{s}um of \underline{s}quared
\underline{d}ifferences between the proportion of positive tests in
$y_{i,j}$ and $\tilde{y}^{(k)}_{i,j}$ (ASSD), computed as
        \begin{eqnarray}
          \tilde{f}_{ASSD}^K(\theta; y) &=& \frac{1}{K} \sum\limits_{k=1}^{K}\sum\limits_{i,j} \left(\frac{y_{i,j,+}}{y_{i,j,\cdot}} - \frac{\tilde{y}^{k}_{i,j,+}}{\tilde{y}_{i,j,\cdot}} \right)^2\,. \label{eq:fASSD}
        \end{eqnarray}
The second is the \underline{s}um of \underline{s}quared of
\underline{d}ifferences between the \underline{a}verage proportion of
positive tests in $y_{i,j}$ and $\tilde{y}^{(k)}_{i,j}$ (SSDA),
computed as
        \begin{eqnarray}
          \tilde{f}_{SSDA}^K(\theta; y) &=& \sum\limits_{i,j} \left(\frac{y_{i,j,+}}{y_{i,j,\cdot}} - \frac{1}{K} \sum\limits_{k=1}^{K} \frac{\tilde{y}^{k}_{i,j,+}}{\tilde{y}_{i,j,\cdot}} \right)^2\,. \label{eq:fSSDA}
        \end{eqnarray}
The third corresponds to the log-likelihood under a \underline{b}inomial model
with \underline{i}ndependent $y_{i,j}$'s (BI), computed as
         \begin{eqnarray}
          \tilde{\ell}_{BI}^K(\theta; y) &=& \sum\limits_{i,j} y_{i,j,+} \log \left(\frac{1}{K} \sum\limits_{k=1}^{K} \frac{\tilde{y}^{k}_{i,j,+}}{\tilde{y}_{i,j,\cdot}} \right) +  y_{i,j,-} \log \left(1 - \frac{1}{K} \sum\limits_{k=1}^{K} \frac{\tilde{y}^{k}_{i,j,+}}{\tilde{y}_{i,j,\cdot}} \right)\,. \label{eq:lBI}
        \end{eqnarray}
The objective functions $f^K_{ASSD}$ and $f^K_{SSDA}$ are to be
minimized while the objective functions $\ell^K_{BI}$ is to be
maximized. Variations around Eq.~\ref{eq:fASSD}-\ref{eq:lBI} are
possible. In particular, the option of replacing the proportion of
positive tests $\tilde{y}^{k}_{i,j,+}/\tilde{y}_{i,j,\cdot}$ by its
expected value (Eq.~\ref{eq:psi}) was considered, and the
corresponding objective functions are referred to as $f^K_{ASSDe}$,
$f^K_{SSDAe}$, and $\ell^K_{BIe}$. For $f^K_{ASSD}$ and $f^K_{SSDA}$,
reweighing each term $(i,j)$ of the sum by $y_{i,j,\cdot}^2$ changes
the differences between proportions into differences between counts
and led to $f^K_{ASSDc}$ and $f^K_{SSDAc}$.

\subsubsection*{Optimization based on Nelder-Mead algorithm}

Evaluation of the objective function is computationally expensive due
to the size of the simulated system (12,857 farms linked through
919,304 animal movements) and the need for repetitions (Monte Carlo
integration with $K=100$ in practice).

Nelder-Mead algorithm (also known as Simplex)~\cite{nelder1965simplex}
was chosen among the multitude of optimization algorithms. This
algorithm has the advantage of being efficient, straightforward to
implement, and derivative-free. To explore a $I$-dimensional space the
algorithm captures information on the local shape of the objective
function by using $I+1$ test points (the simplex). Iterations use this
information to update the simplex by choosing between reflection,
expansion, contraction, and shrink steps. Given the usual formulation
of the algorithm as solving a minimization problem, we minimized
$-{\ell}^K_{CC}(\theta; y)$. Also, since the algorithm works in
$\mathbb{R}^I$, to restrain the optimization to the parameter space
defined by the minimum and maximum values allowed for each parameter
($\theta_i \in ]\theta_i^{min}, \theta_i^{max}[$), logit
    transformations were applied (optimization on $x_i \in
  ]-\infty,+\infty[$ and $\theta_i=\theta_i^{min} +
    (\theta_i^{max}-\theta_i^{min})/(1+e^{-x_i})$).

To cope with stochastic noise in the evaluation of the objective
function, the Nelder-Mead algorithm was adapted along the lines
proposed by \cite{BartonRussell1991}. These adaptations consist of (i)
re-evaluation of the objective function at all points of the simplex
after a shrink step, (ii) choice of a small amplitude for the shrink
step (coefficient of 0.9 instead of the more commonly used value of
0.5), and (iii) restarting several times the algorithm by constructing
a fresh simplex with the size of the initial regular simplex, centered
on the current best point. When not specified otherwise in the text,
such fresh restarts are done every 100 iterations.

Starting from an initial simplex, the output of the algorithm is a
sequence of $(\theta^{(m)})_{m=1\ldots M}$ whose each element is
associated with an evaluation of the objective function,
$\tilde{\ell}^K_{CC}(\theta^{(m)}; y)$. The
$\tilde{\ell}^K_{CC}(\theta^{(m)}; y)$ are not always increasing due to
the adaptations (i) and (iii). The final point estimate $\hat{\theta}$
was chosen as $\theta^{(m^\star)}$ with $m^\star = \arg\max_m
\tilde{\ell}^K_{CC}(\theta^{(m)}; y)$. An estimate of
$\ell^K_{CC}(\hat{\theta}; y)$, denoted
$\bar{\ell}^K_{CC}(\hat{\theta}; y)$, was obtained by averaging $10$ independent
evaluations of $\tilde{\ell}^K_{CC}(\hat{\theta}; y)$.

\subsubsection*{From point estimates to confidence intervals using profiles}

Use of Nelder-Mead algorithm leads to a point estimate $\hat{\theta}$
of the parameters in a space of dimension $I=7$ and an estimate of the
corresponding value of the maximized objective function
$\bar{\ell}^K_{CC}(\hat{\theta}; y)$. A "profile likelihood" approach
served to obtain confidence intervals for each parameter $i$. Namely,
confidence intervals at level $\alpha$ are defined for the
$i$\textsuperscript{th} parameter as the range of values for
$\theta_i$ which are compatible with a value of the objective function
above $\bar{\ell}^K_{CC}(\hat{\theta}; y) - \tau_\alpha$. In practice,
these intervals are delineated after establishing a profile of
$\max_{\theta_{-i}}\bar{\ell}^K_{CC}((\theta_i,\theta_{-i}); y)$ as a
function of $\theta_i$, where $\theta_{-i}$ denotes all the parameters
but $\theta_i$. This is achieved by fixing $\theta_i$ to various
values and performing maximization with respect to $\theta_{-i}$ ({\sl
  i.e.} in a space of dimension $I-1$) using Nelder-Mead
algorithm. For this purpose, the initial simplex is chosen centered on
$\hat{\theta}_{-i}$ and periodic restarts are deactivated.

Whereas in the classical maximum likelihood inference framework
asymptotic results justify the use of quantiles of chi-squared
distribution \cite{venzon1988method}, the tolerance threshold
$\tau_\alpha$ for level $\alpha=95\%$ needed here to be calibrated by
estimation on simulated datasets (see results).

\section*{Results}

\subsection*{Choice of the objective function on simulated datasets}

An extensive simulation study was conducted to compare the
performances of different candidate objective functions that could be
used for inference. This numerical study started by selecting 2,000
scenarios, each corresponding to a distinct value of
$\theta$. Dispersion of the scenarios and hence coverage of the
parameter space was maximized by Latin Hypercube Sampling (LHS),
accounting for a constraint on the shape of the within-herd prevalence
distribution: $\Phi_{\theta_{mi},\theta_{si}}(0) >= 0.1$, in order to
avoid too many similar truncated distribution shapes.
For each scenario, the simulation model
was used to generate 101 epidemiological trajectories, out of which 1
served as an observed dataset (i.e. playing the role of real
data). The other 100 served to compute the terms of the objective
function needed to compare any observed data to this particular
scenario using any objective function (i.e. playing the role of the
simulated data with $K=100$ in
Eq.~\ref{eq:approxconditionalcompositelikelihood} and
\ref{eq:approxcompositelikelihood}). Each simulated observed dataset
was then compared to each scenario (2,000 $\times$ 2,000 comparisons)
using 16 candidate objective functions.

Performance of the 5 main types of objective functions, in terms of
ability to identify the scenario that generated the observed data, are
reported in Table~\ref{tab:comparisoncriteriaperf} (see
\nameref{S1Table} for the other candidate objective
functions). Objective functions based on approximate composite
likelihoods ($\tilde{\ell}^{100}_{MC}(\theta; y)$ and
$\tilde{\ell}^{100}_{CC}(\theta; y)$) were able to identify the
correct scenario (among 2,000) for more than 95\% of the simulated
observed datasets. These functions produced about twice less errors
than more traditional objective functions based on the sum of squared
deviations from the mean of the 100 simulations
($\tilde{f}^{100}_{SSDA}$ and $\tilde{f}^{100}_{SSDAc}$) which
achieved about 90\% of correct scenario identification. Objective
functions built as the sum of squared differences between observed
data and each of the 100 simulations before averaging the results
($\tilde{f}^{100}_{ASSD}$ and $\tilde{f}^{100}_{ASSDc}$) obtained
dramatically lower performances with only about 2\% of correct
scenario identification.

Markov transition terms introduced in $\tilde{\ell}^{100}_{CC}(\theta;
y)$ to take into account dependence between consecutive sampling
points in the same herd did not produce benefits that could be easily
seen Table~\ref{tab:comparisoncriteriaperf} (in the comparison with
$\tilde{\ell}^{100}_{MC}(\theta; y)$). Another perspective for the
comparison of $\tilde{\ell}^{100}_{CC}(\theta; y)$ and
$\tilde{\ell}^{100}_{MC}(\theta; y)$, is to see these approximate
composite likelihoods as regular likelihoods associated with two
different models, each one attempting to approach the true
distribution of the observed data. A closer representation of the true
model is thus expected to lead to a higher value of the approximate
composite likelihood. From this perspective,
$\tilde{\ell}^{100}_{CC}(\theta; y)$ is clearly better than
$\tilde{\ell}^{100}_{MC}(\theta; y)$ since
$\tilde{\ell}^{100}_{CC}(\theta^0; y)$ was higher than
$\tilde{\ell}^{100}_{MC}(\theta^0; y)$ for $99.55\%$ of the couples
$(\theta^0,y)$ where $\theta^0$ denotes the parameter values that
served to generate $y$ (\nameref{S1Fig}). The
approximate composite likelihood $\tilde{\ell}^{100}_{CC}(\theta; y)$
was thus selected as the objective function maximized with respect to
$\theta$ for inference.

\begin{table}[!t]
\begin{adjustwidth}{-2.25in}{0in} 
        \setlength\extrarowheight{2pt}
\centering
\caption{
{\bf Evaluation of candidate objective functions for inference based on a LHS of 2,000 scenarios.}}
\begin{tabular}{lccccc}
\thickhline
                                                &  \multicolumn{5}{c}{Objective function\textsuperscript{b}}    \\
 Performance measure\textsuperscript{a}                            &  $\tilde{f}^{100}_{ASSD}$  &  $\tilde{f}^{100}_{SSDA}$  &  $\tilde{\ell}^{100}_{BI}$    &  $\tilde{\ell}^{100}_{MC}$  &  $\tilde{\ell}^{100}_{CC}$  \\
 \thickhline
\multicolumn{6}{l}{Rank of $\theta^0$} \\
\hspace{3cm}first ({\it i.e.} $\theta^* = \theta^0$)  &  0.170    &  0.897  &   0.754  & \cellcolor{Gray!40} 0.970  &  \cellcolor{Gray!20} 0.954   \\
\hspace{3cm}top 5\%                                  &  0.369   &  \cellcolor{Gray!40} 1.00    &  \cellcolor{Gray!40} 1.00    &  \cellcolor{Gray!40} 1.00    &  \cellcolor{Gray!40} 1.00    \\
\hline
\multicolumn{6}{l}{Error on parameter ($|\theta^{*}_{\cdot}-\theta^0_{\cdot}|/|\theta^{\max}_{\cdot}-\theta^{\min}_{\cdot}|$)} \\
\hspace{3cm}$\theta_{hi}$        &  0.285  &  0.019  &  0.034  &  \cellcolor{Gray!40} 0.001  &  \cellcolor{Gray!20} 0.002  \\
\hspace{3cm}$\theta_{mi}$      &  0.259  &  0.027  &  0.072  &  \cellcolor{Gray!40} 0.006  &  \cellcolor{Gray!20} 0.011  \\
\hspace{3cm}$\theta_{si}$        &  0.352  &  0.027  &  0.069  &  \cellcolor{Gray!40} 0.006  &  \cellcolor{Gray!20} 0.009  \\
\hspace{3cm}$\theta_{mi^t}$      &  0.206  &  0.017  &  0.043  &  \cellcolor{Gray!40} 0.004  & \cellcolor{Gray!20} 0.006  \\
\hspace{3cm}$\theta_{si^t}$        &  0.278  &  0.019  &  0.048  &  \cellcolor{Gray!40} 0.004  & \cellcolor{Gray!20} 0.007 \\
\hspace{3cm}$\theta_{oi}$      &  0.278  &  0.015  &  0.040  &  \cellcolor{Gray!40} 0.004  &  \cellcolor{Gray!20} 0.005  \\
\hspace{3cm}$\theta_{od}$    &  0.254  &  0.016  &  0.039  &  \cellcolor{Gray!40} 0.005  &  \cellcolor{Gray!20} 0.007  \\
\hspace{3cm}$\theta_{bg}$        &  0.146  &  0.008  &  0.023  &  \cellcolor{Gray!40} 0.002  &  \cellcolor{Gray!20} 0.003  \\
\hspace{3cm}$\theta_{se}$  &  0.196  &  0.012  &  0.022  &  \cellcolor{Gray!40} 0.002  &  \cellcolor{Gray!20} 0.003  \\
\hline
\multicolumn{6}{l}{Error on parameter, after excluding the true scenario ($|\theta^{+}_{\cdot}-\theta^0_{\cdot}|/|\theta^{\max}_{\cdot}-\theta^{\min}_{\cdot}|$)} \\
\hspace{3cm}$\theta_{hi}$        &  0.287  &  0.157  &  0.152  &  \cellcolor{Gray!40} 0.080  &  \cellcolor{Gray!20} 0.085  \\
\hspace{3cm}$\theta_{mi}$       &  0.262  &  0.276  &  0.282  &  \cellcolor{Gray!20} 0.246  &  \cellcolor{Gray!40} 0.245  \\
\hspace{3cm}$\theta_{si}$        &  0.358  &  0.288  &  0.299  &  \cellcolor{Gray!20} 0.257  &  \cellcolor{Gray!40} 0.253  \\
\hspace{3cm}$\theta_{mi^t}$       &  0.209  &  0.171  &  0.173  &  \cellcolor{Gray!20}  0.143  & \cellcolor{Gray!40} 0.139  \\
\hspace{3cm}$\theta_{si^t}$       &  0.283  &  0.211  &  0.216  &  \cellcolor{Gray!20}  0.174  & \cellcolor{Gray!40} 0.170    \\
\hspace{3cm}$\theta_{oi}$        &  0.281  &  0.166  &  0.173  &  \cellcolor{Gray!20} 0.148  &  \cellcolor{Gray!40} 0.148  \\
\hspace{3cm}$\theta_{od}$        &  0.258  &  0.177  &  0.178  &  \cellcolor{Gray!20} 0.175  &  \cellcolor{Gray!40} 0.171  \\
\hspace{3cm}$\theta_{bg}$        &  0.148  &  \cellcolor{Gray!40} 0.076  &  0.084  & 0.081  &   \cellcolor{Gray!20} 0.077  \\
\hspace{3cm}$\theta_{se}$        &  0.197  &  0.083  &  0.082  &  \cellcolor{Gray!20} 0.059  &  \cellcolor{Gray!40} 0.057  \\
\thickhline
\end{tabular}
\begin{flushleft} \textsuperscript{a} For each performance measure, results reported here are average over 2,000 simulated observed datasets. Notations: $\theta^0$, the set of parameters that served to simulate the observed dataset ({\it i.e.} the ``thruth''); $\theta^\star$, the set of parameters that maximized the objective function, and $\theta^+$, the set of parameters that maximized the objective function after excluding $\theta^0$. \textsuperscript{b} Objective functions described in the text, identified by subscript labels: $ASSD$, average sum of squared differences; $SSDA$, sum of squared of differences; $BI$, binomial independent; $MC$, marginal composite likelihood; $CC$, conditional composite likelihood. Gray and light gray backgrounds highlight, respectively, best and second best objective functions with respect to each performance measure.
\end{flushleft}
\label{tab:comparisoncriteriaperf}
\end{adjustwidth}
\end{table}

\subsection*{Validation of point estimates on simulated datasets}

    \begin{figure}[!t]
    \centering 
    \includegraphics[width=0.99\textwidth]{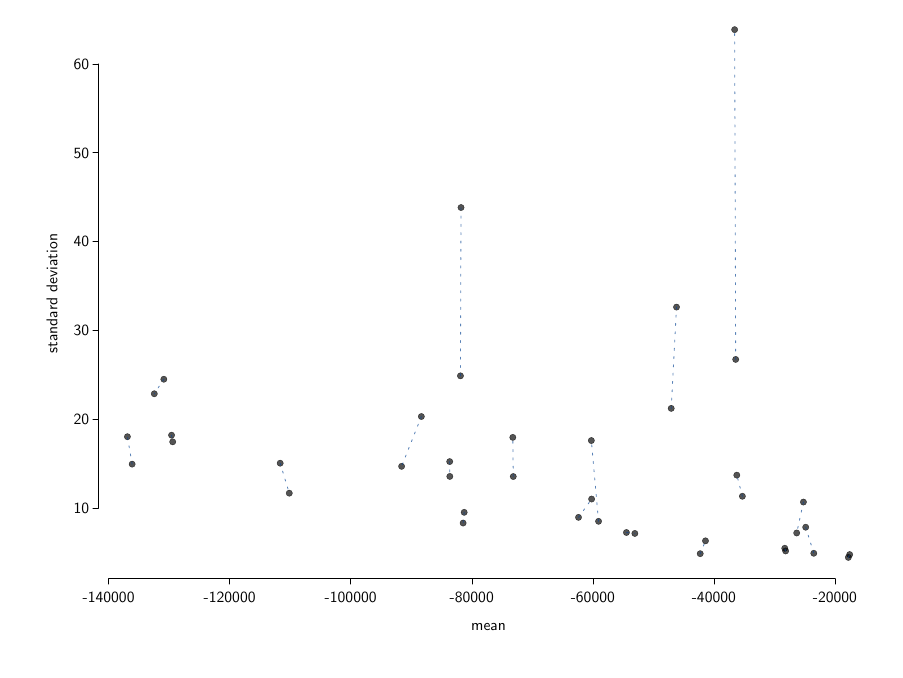}
    \caption{Mean and standard deviation of the approximate composite
      likelihood on simulated observed datasets. For each of the 40
      simulated observed dataset $y$, the mean (x-axis) and standard
      deviation (y-axis) of 20 evaluations of
      $\tilde{\ell}^{100}_{CC}(\theta^0;y)$ are reported, where
      $\theta^0$ is the set of parameters that served to generate
      $y$.}
    \label{fig:mean_and_sd_of_objectivefunction}
    \end{figure}

\begin{figure}[!t]
  \centering
  \includegraphics[width=0.99\textwidth]{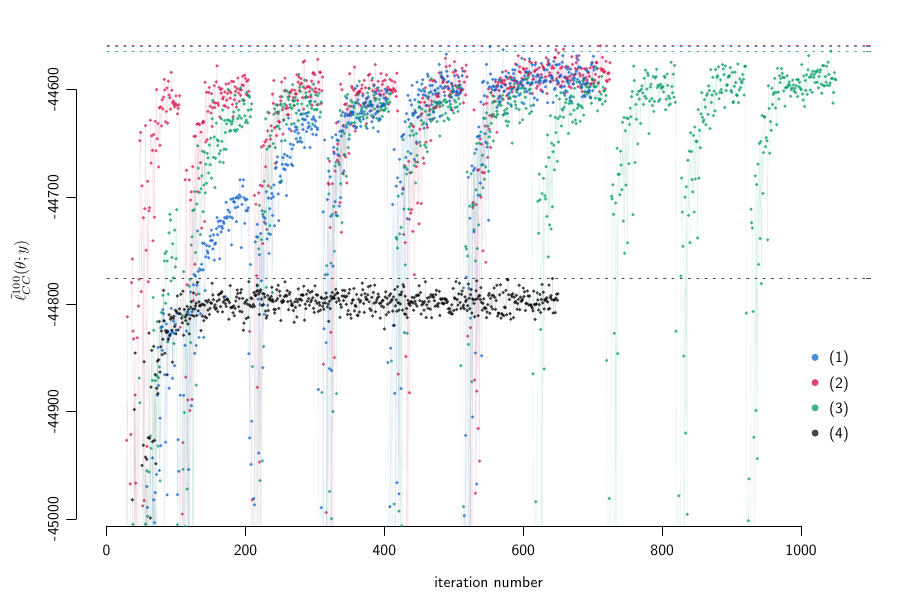}
  \caption{Behaviour of the optimization procedure used for parameter
    estimation.  Iterations of the modified Nelder-Mead algorithm are
    counted in number of evaluations of the approximate composite
    likelihood $\tilde{\ell}^{100}_{CC}(\theta; y)$
    used as objective function. Four trajectories obtained
    on the real observed dataset from Brittany are represented with
    different colors. Blue, red, and green trajectories differ by
    their starting points while the black one corresponds to a trajectory
    without periodic fresh restarts restoring the initial size of the
    simplex.}
  \label{fig:simplextrajectories}
\end{figure}

    \begin{figure}[!t]
    \begin{adjustwidth}{-2.25in}{0in} 
    \centering 
    \includegraphics[angle=0,width=0.85\paperwidth]{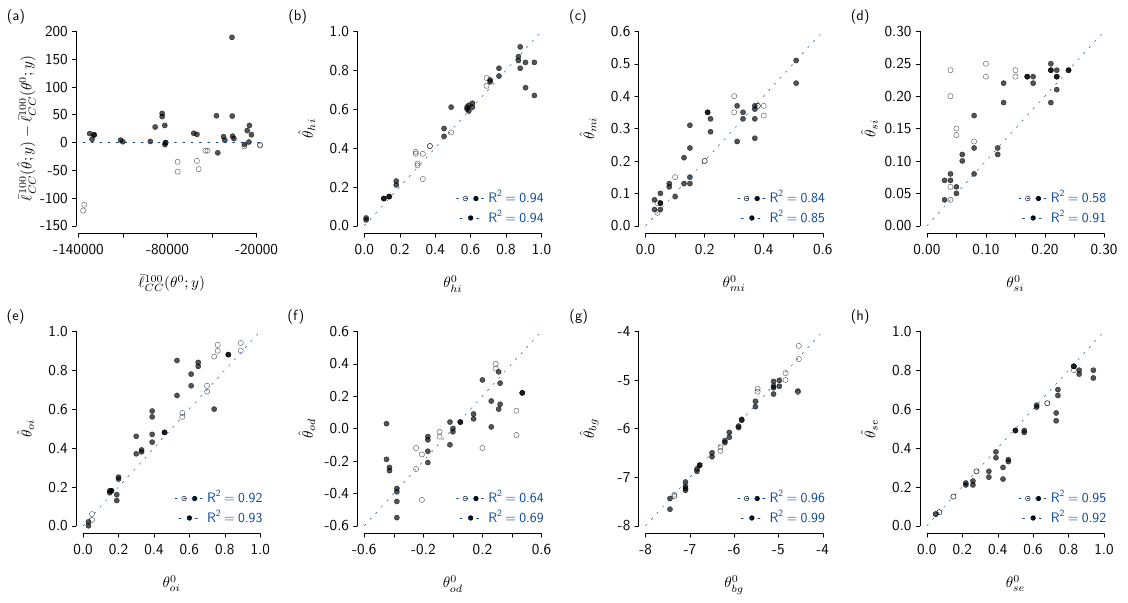}
    \caption{Characteristics of point estimates of the parameters on
      simulated datasets. Results are based on independent point
      estimations carried out on 40 distinct datasets. (a) Comparison
      between the value of the approximate composite likelihood
      $\bar{\ell}^{100}_{CC}(\theta;y)$ evaluated at $\hat{\theta}$
      (point estimate) and $\theta^0$ (set of parameters used to
      simulate the observed data). Datasets are represented by black
      points when $\bar{\ell}^{100}_{CC}(\hat{\theta};y) >
      \bar{\ell}^{100}_{CC}(\theta^0;y)$ or when the difference is not
      statistically different at the 5\% significance level as
      assessed with a t-test comparing the 10 evaluations of
      $\tilde{\ell}^{100}_{CC}(\theta;y)$ used to compute
      $\bar{\ell}^{100}_{CC}(\theta;y)$ for each $\theta$. White
      points are used otherwise. (b-h) Comparison between
      $\hat{\theta}$ and $\theta^0$ for each of the seven
      model parameters; same color code as in (a).}
    \label{fig:pointestimates_on_simulateddata}
    \end{adjustwidth} 
    \end{figure}

The estimation procedure based on numerical optimization was evaluated
on a set of 20 scenarios selected by a LHS. Here,
the number of scenarios was lower than analyzed to choose the
objective function due to the computational cost of the optimization,
which requires many evaluations of the objective function. In
practice, a total of 40 observed datasets were simulated, with 2
independent simulations for each of the 20 scenarios to unveil
variability associated with the randomness of the observed dataset
given the parameter values.

These 40 simulated observed datasets from 20 different scenarios
covered a wide range of values for the mean and standard deviation of
the objective function (the scenarios were constructed on the basis of
two LHS, including one for which the parameters of the within-herd
prevalence distribution were manually defined).
As shown in Fig.~\ref{fig:mean_and_sd_of_objectivefunction}, the
approximate composite log-likelihood $\tilde{\ell}^{100}_{CC}(\theta^0;y)$
varied across the 20 scenarios by an amplitude of approximately 7-fold
in terms of estimated mean (from around -140,000 to -20,000) and more
than 10-fold in terms of estimated standard deviation (from around 5 to 60).
Not unexpectedly, Fig.~\ref{fig:mean_and_sd_of_objectivefunction} shows a
trend for a positive correlation between the absolute value of the mean and
the standard deviation and some differences between duplicated simulations.

For each dataset, numerical maximization of
$\ell^{100}_{CC}(\theta;y)$ with respect to $\theta$ lead to a point
estimate $\hat{\theta}$. Fig.~\ref{fig:simplextrajectories} illustrates
the behaviour of this numerical optimization on the
real data and demonstrates the importance of periodic fresh restarts to
minimize the impact of the starting point.
Fig.~\ref{fig:pointestimates_on_simulateddata} shows the results
obtained for the 40 simulated datasets based on 3 fresh restarts (300
iterations). Attesting of the effectiveness of the optimization
procedure, point estimates $\hat{\theta}$ were always such as
$\bar{\ell}^{100}_{CC}(\hat{\theta};y)$ is close to
$\bar{\ell}^{100}_{CC}(\theta^0;y)$
(Fig.~\ref{fig:pointestimates_on_simulateddata}a). Of note, an ideal
optimization procedure is expected to reach a value for
$\bar{\ell}^{100}_{CC}(\hat{\theta};y)$ higher than
$\bar{\ell}^{100}_{CC}(\theta^0;y)$. This was the case for 26 (65\%)
of the simulated datasets.

Comparison of point estimates, $\hat{\theta}$, with true values of the
parameters used to simulate the data, $\theta^0$, revealed a good
performance of the estimators, as reported in
Fig.~\ref{fig:pointestimates_on_simulateddata}b-h (details in
\nameref{S2Table}). For 4 out of the 7 parameters, the fraction of
explained variance ($R^2$) was even higher than 0.9. The 3 parameters
that did not reach this threshold were $\theta_{si}$, $\theta_{od}$,
and $\theta_{mi}$, with $R^2$ respectively of 0.58, 0.64 and 0.84. Two
distinct factors could contribute to lower $R^2$. The first is genuine
statistical uncertainty of point estimates which would translate into
a relatively flatter profile of the objective function and larger
confidence intervals (see below). This uncertainty could be due to a
smaller impact of the parameter on the model outcome or to the
possibility of compensation between parameters. In particular,
interplay between $\theta_{mi}$ and $\theta_{si}$ in the definition of
the initial distribution of prevalence in infected herds may pose
difficulties for point estimations (\nameref{S2Fig}). The second
factor is sub-optimal numerical maximization of the objective
function. Suggesting that sub-optimal maximization plays a role in the
lower $R^2$, we noticed that for the 3 parameters with $R^2<0.9$ the
$R^2$ could be increased by considering only the datasets for which
$\bar{\ell}^{100}_{CC}(\hat{\theta};y)>\bar{\ell}^{100}_{CC}(\theta^0;y)$
(see Fig.~\ref{fig:pointestimates_on_simulateddata}b-h). The effect
was particularly strong for $\theta_{si}$, for which $R^2$ raised from
0.58 to 0.90. Of note, the two factors contributing to lower $R^2$ are
not expected to be disconnected since maximization is more difficult
when the objective function is not steep compared to the level of
noise in its evaluation.

    \begin{figure}[!t]
    \centering 
    \includegraphics[width=0.99\textwidth]{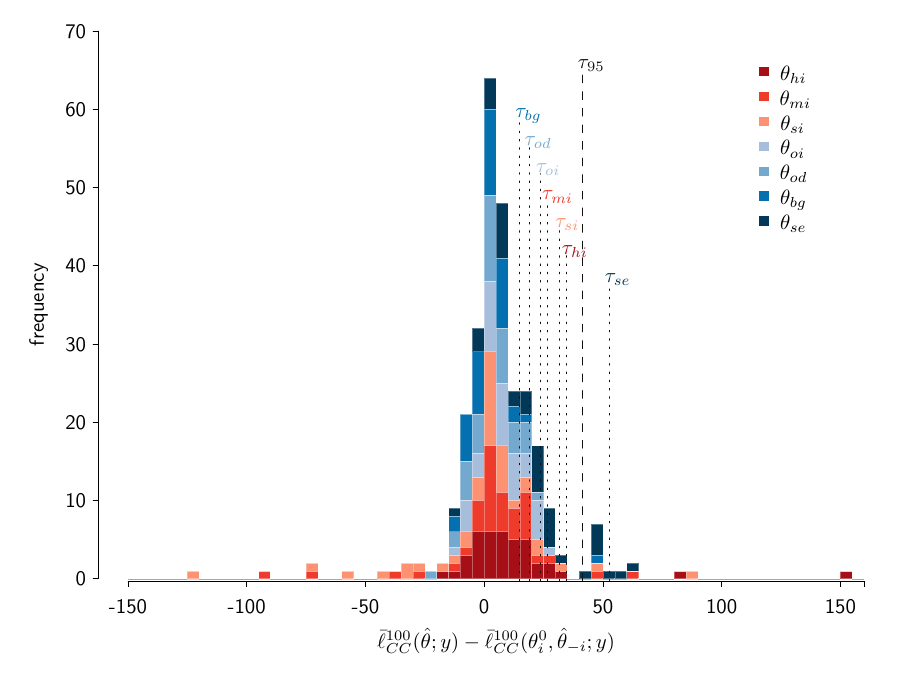}
    \caption{Calibration of thresholds to build confidence
      intervals. Empirical distributions of
      $\bar{\ell}^{100}_{CC}(\hat{\theta};y)-\bar{\ell}^{100}_{CC}(\theta_i^0,\hat{\theta}_{-i};y)$
      are represented along with quantiles at level $95\%$.
      Colors distinguish the 7 parameters of the
      model. Quantiles of the total aggregated distribution and
      specific of each parameter are reported.}
    \label{fig:calibration_threshold}
    \end{figure}

    \subsection*{Calibration of thresholds to build confidence intervals}

    Confidence intervals are built by delineation of regions in the
    parameter space that are close, in terms of value of the objective
    function, to $\hat{\theta}$ which represents the best set of
    parameter values found for the dataset. Under the conditions of
    the asymptotic theory of maximum likelihood estimation, thresholds
    $\tau_\alpha$ that define confidence intervals of one-dimensional
    parameters are given by the quantiles of the Chi-squared
    distribution with one-degree of freedom. Boundaries of intervals
    at $\alpha=95\%$ are then located where the value of the
    log-likelihood drops $1.92$ below its maximum. By using an
    objective function built as an approximate composite
    log-likelihood, we are not in the conditions of this theory, and
    we thus conducted a numerical analysis to calibrate the thresholds
    that could serve to build confidence intervals.

    Thresholds corresponding to a target confidence level of $95\%$
    were calibrated on the 40 simulated datasets that served to
    validate point estimates. For each simulated dataset and each
    parameter $i$ (out of the 7 model parameters), we examined the
    distribution of the difference
    $\bar{\ell}^{100}_{CC}(\hat{\theta};y)-\bar{\ell}^{100}_{CC}(\theta_i^0,\hat{\theta}_{-i};y)$,
    where $\hat{\theta}_{-i}$ is obtained by numerical optimization
    with $\theta_i$ held fixed at $\theta_i^0$. Based on the histogram
    aggregating all datasets and all parameters
    (Fig.~\ref{fig:calibration_threshold}), the threshold was
    established at $41.18$.

    The empirical distribution shown in
    Fig.~\ref{fig:calibration_threshold} is far from a Chi-squared
    distribution with one-degree of freedom. This was anticipated not
    only given the use of an objective which is not the genuine
    log-likelihood of the model that generated the data, but also
    given noisy evaluation of the objective function and the
    associated sub-optimal behavior of the numerical optimization. To
    assess robustness with respect to this last point we examined if
    the distribution differed when excluding the datasets for which
    the point estimate is farther from its optimal value as detected
    by $\bar{\ell}^{100}_{CC}(\hat{\theta};y)$ being significantly
    below $\bar{\ell}^{100}_{CC}(\theta^0;y)$ (white points in
    Fig.~\ref{fig:pointestimates_on_simulateddata}). No clear impact
    of these points was observed since the resulting threshold was
    established at $45.12$ (\nameref{S3Fig}). Importantly, departure
    from the conditions of the asymptotic theory of maximum likelihood
    estimation makes that the hypothesis of a common threshold for all
    parameters may not hold. To cope with this issue, thresholds were
    also established separately for each parameter
    (Fig.~\ref{fig:calibration_threshold}(b)). This confirmed a
    substantial variability between parameters, with thresholds
    ranging from 14.83 (for $\theta_{bg}$) to 52.45 (for
    $\theta_{se}$). For all but $\theta_{se}$ these parameter-specific
    thresholds were lower than the global threshold. Their use tends
    thus to lead to narrower, more optimistic, confidence
    intervals. To be conservative, the two types of confidence
    intervales, based on global and parameter-specific thresholds, are
    reported for real data below.

    \subsection*{Estimation of epidemiological parameters on serological data from Brittany}

    To circumvent the risk of sub-obtimal point estimation due to the
    difficulty of maximizing the objective function, the results of
    numerical optimization from three distant starting points were
    compared and extra iterations of the algorithm were performed
    (Fig.~\ref{fig:simplextrajectories}). A first starting point was
    the center of the parameter space (center of the interval of
    allowed values for each parameter), a second starting point was
    based on experts opinion (potentially close to probable values), a
    third starting point was selected to be distant from expert
    opinion. After following different paths in the parameter
    space (\nameref{S4Fig}), the three numerical optimization trajectories
    converged towards similar values of the objective
    function and of the parameters (Table~\ref{tab:valparamestim}). It
    is therefore very likely that the region of the parameter space
    where the objective function is maximum has been correctly
    identified.

\begin{figure}[!t]
    \begin{adjustwidth}{-2.25in}{0in} 
    \centering 
    \includegraphics[width=0.85\paperwidth]{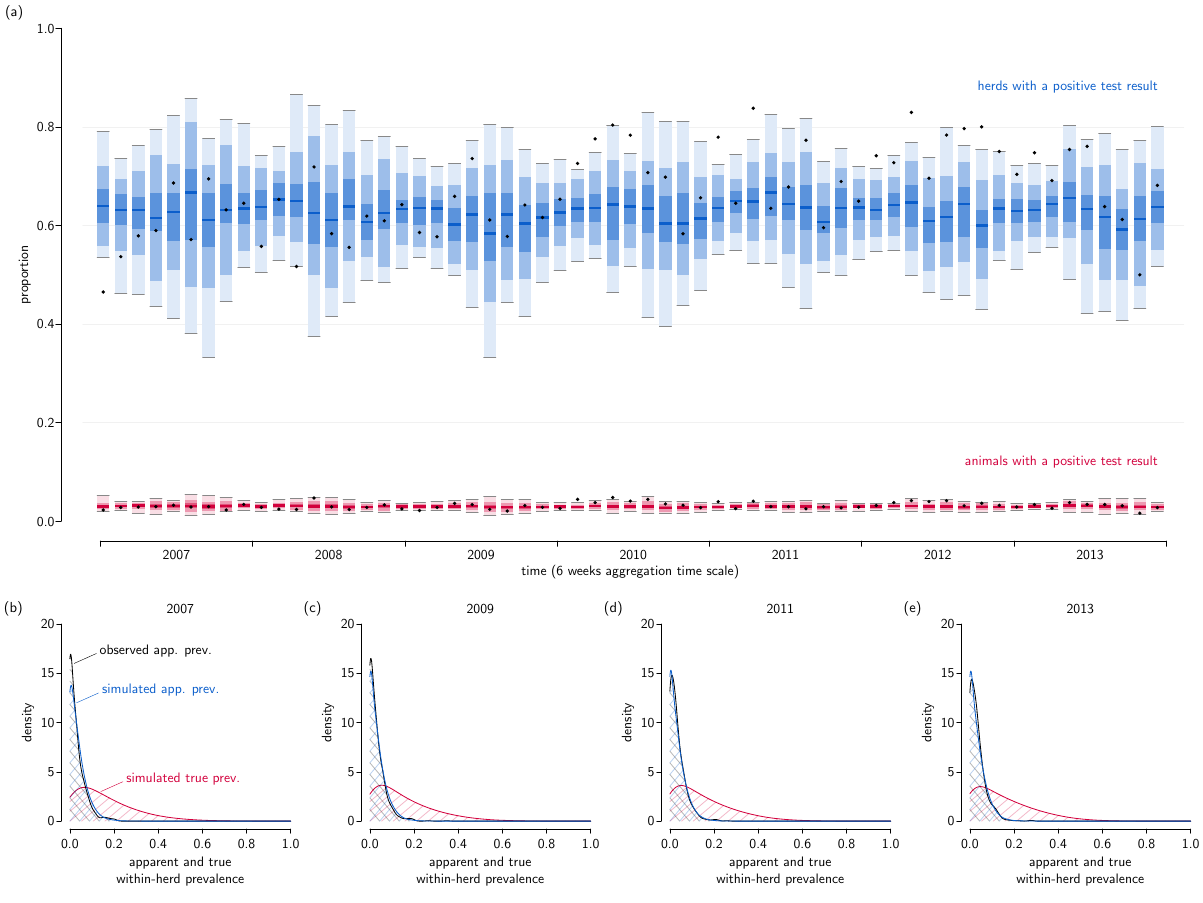}
    \caption{Observation versus simulation. (a) proportion of herds (in blue) and
    animals (in red) with a positive test result. Black points correspond to
    the observed data. For the simulation, each bar is divided into four
    levels of color intensity, corresponding, from darkest to lightest,
    to the median, and to the quantiles 25-75, 5-95 and 0-100, computed
    from a simulation of 200 runs with the estimated parameter
    values. (b - e) within-herd prevalence per year. Distributions
    in black and blue, correspond, respectively, to the observed data
    and simulated predictions of the apparent within-herd prevalence,
    and distributions in red correspond to the simulated predictions
    of the true within-herd prevalence, among the sampled herds, for
    a given year. }
    \label{fig:obs_vs_simu}
    \end{adjustwidth} 
    \end{figure}

    To rule out the possibility of clear discrepancy between the model
    and the data that would jeopardize the relevance of the estimates,
    we compared epidemiological trajectories simulated with parameters
    corresponding to these point estimates to the real data. For this
    purpose, the distribution of positive tests over time in the
    observed data was examined both in terms of fraction of herd
    screening events that reported at least a positive test
    (Fig.~\ref{fig:obs_vs_simu}a) and in terms of distribution of the
    fraction of positive tests in screening events
    (Fig.~\ref{fig:obs_vs_simu}b-e). The simulated trajectories
    reproduced faithfully these characteristics of the real data.

    Point estimates and their associated confidence intervals allow to
    draw a detailed picture of the status and dynamics of Map spread
    in Brittany. The key parameter that connects observed and real
    prevalence of the disease is $\theta_{se}$, the sensitivity of the
    diagnostic test. This sensitivity is estimated at $0.21$. A
    total of 5 parameters characterize the prevalence at the temporal
    and spatial boundaries of the metapopulation. For the temporal
    aspect, the initial prevalence is characterized by a very high
    proportion of initially infected herds
    ($\hat{\theta}_{\mathit{hi}} = 0.97$) but a relatively low
    prevalence within these herds (mean $\hat{\theta}_{mi^t}=0.17$ and
    standard deviation $\hat{\theta}_{si^t}=0.15$). The resulting
    initial overall prevalence in the metapopulation is $0.16$
    ($\hat{\theta}_{\mathit{hi}}\times \hat{\theta}_{mi^t}$). For the
    spatial aspect, the estimated probability to purchase an infected
    animal from outside the metapopulation appears to be comparable
    to this overall prevalence ($\theta_{\mathit{oi}} =0.14$) and
    almost constant over time ($\theta_{\mathit{od}} \approx
    0$). Finally, the key parameter of the dynamics within the
    metapopulation (animal movements being given) is $\theta_{bg}$,
    the logarithm of the coefficient governing the within-herd
    transmission rate through the general environment of the farm.
    It is estimated to $-6.80$.

\begin{figure}[!t]
    \begin{adjustwidth}{-2.25in}{0in} 
    \centering 
    \includegraphics[angle=0,width=0.85\paperwidth]{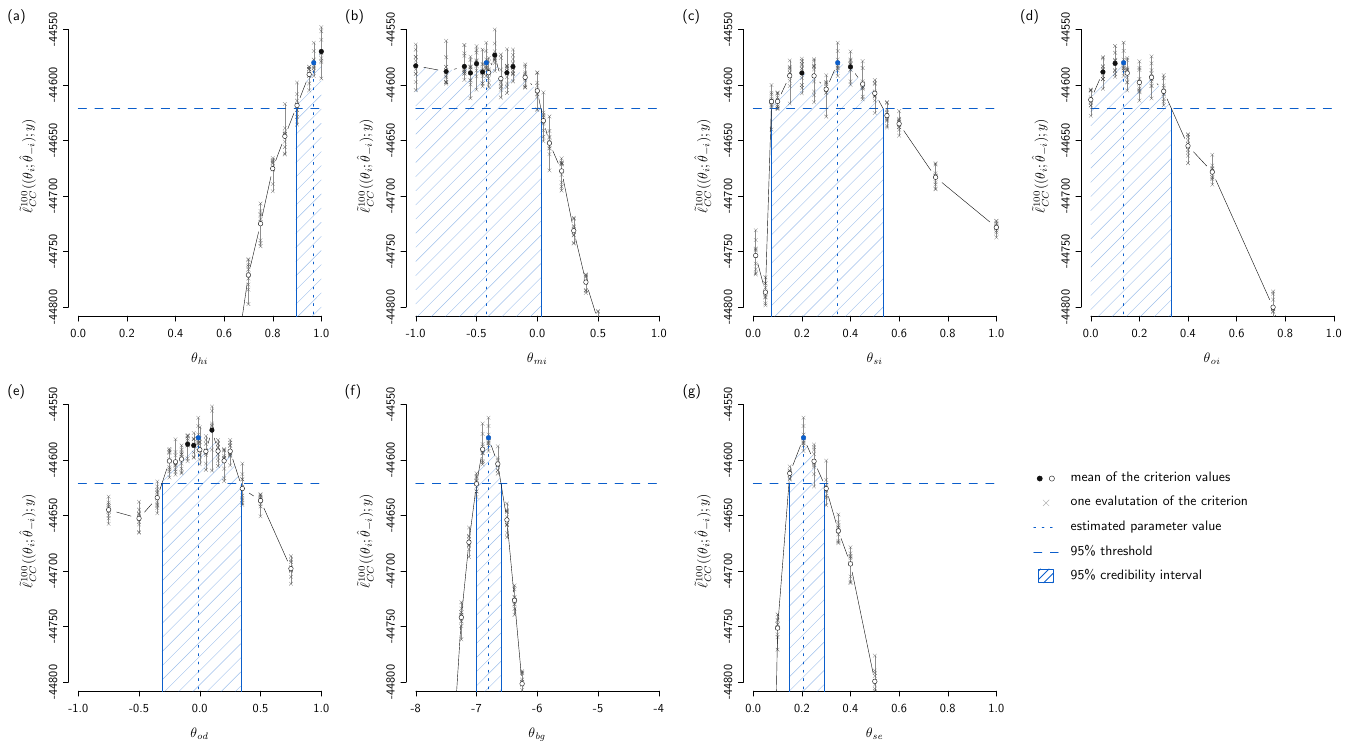}
    \caption{Confidence intervals. Profile likelihood were drawn using local estimations by fixing, one at a time, the parameters at different values. The 95\% confidence intervals are built using corresponding threshold defined before. With black and white points: black when the distribution of $\tilde{\ell}^{100}_{CC}((\theta_{i};\hat{\theta}_{-i});y)$ is not statistically different from the 10 evaluations of $\tilde{\ell}^{100}_{CC}(\hat{\theta};y)$, at the 5\% significance level as assessed with a t-test, white otherwise.}
    \label{fig:profilelikelihood}
    \end{adjustwidth} 
    \end{figure}

    Confidence intervals associated with each of the parameter
    estimates were determined by drawing profiles of the objective
    function and applying the thresholds numerically established on
    simulated data (Fig~\ref{fig:profilelikelihood}). Profiles
    obtained for parameters $\theta_{\mathit{hi}}$,
    $\theta_{\mathit{bg}}$, and $\theta_{\mathit{se}}$ were highly
    discriminating, with large differences in the objective function
    across the range of possible parameter values. Accordingly,
    confidence intervals (Table~\ref{tab:valparamestim}) established
    for these parameters are narrower than for the initial
    distribution of within-herd prevalence ($\theta_{\mathit{mi}}$ and
    $\theta_{\mathit{si}}$) and for the probability of purchasing an
    infected animal from outside the metapopulation
    ($\theta_{\mathit{oi}}$ and $\theta_{\mathit{od}}$). The
    estimations conducted in order to draw the profiles show clear
    compensation effects between some parameters
    (\nameref{S5Fig}).
    As mentioned above, the parameters of the initial within-herd
    prevalence distribution ($\theta_{\mathit{mi}}$ and
    $\theta_{\mathit{si}}$) evolved together to obtain similar
    truncated distribution shapes, leading to a relatively flat
    profile for the objective function near their point estimates.
    For the probability of purchasing infected cattle from outside the
    metapopulation ($\theta_{\mathit{oi}}$ and
    $\theta_{\mathit{od}}$), we also observed a negative correlation
    between intercept and slope, with $\theta_{od}>0$ when
    $\theta_{oi}<\hat{\theta}_{oi}$ and $\theta_{od}<0$ when
    $\theta_{oi}>\hat{\theta}_{oi}$.

    \begin{table}[t]
      \begin{adjustwidth}{-2.25in}{0in} 
        \setlength\extrarowheight{2pt}
    \centering 
    \caption{
    {\bf Estimated parameter values on real epidemiological survey data (Brittany, France)}}
    \begin{tabular}{@{\extracolsep{6pt}}lrrrrrrr}
      \thickhline
      & & & & \multicolumn{4}{c}{Confidence intervals (95\%)}\\
      \cline{5-8}
      & \multicolumn{3}{c}{Point estimates\textsuperscript{a}} & \multicolumn{2}{c}{shared $\tau_{95}$\textsuperscript{b}} & \multicolumn{2}{c}{param.-sp. $\tau_{95}$\textsuperscript{c}} \\
      \cline{2-4}
      \cline{5-6}
      \cline{7-8}
      &  Run 1 & Run 2 & Run 3 & Min & Max & Min & Max \\
      \thickhline
    $\theta_{\mathit{hi}}$  &  \cellcolor{Gray!20} 0.97 & 0.97 & 0.95     & 0.90 &   $\overline{1.00}$ & 0.91 &   $\overline{1.00}$ \\
    $\theta_{\mathit{mi}}$ ($\theta_{\mathit{mi^t}}$) &  \cellcolor{Gray!20} -0.42 (0.17) & -0.37 (0.17) & -0.43 (0.18)  & $\underline{-1.00}$ &  0.03 & $\underline{-1.00}$ &  0.00 \\
    $\theta_{\mathit{si}}$ ($\theta_{\mathit{si^t}}$) &  \cellcolor{Gray!20} 0.35 (0.15)  & 0.34 (0.14)  & 0.37 (0.16)  & 0.07 &  0.53 & 0.11 &  0.51 \\
    $\theta_{\mathit{oi}}$  &   \cellcolor{Gray!20} 0.14 & 0.15 & 0.13    & $\underline{0.00}$ &  0.33 & 0.02 &  0.29 \\
    $\theta_{\mathit{od}}$  &  \cellcolor{Gray!20} -0.01 & 0.03 & 0.02   & -0.31 &  0.34 & -0.15 &  0.27 \\
    $\theta_{\mathit{bg}}$  &  \cellcolor{Gray!20} -6.80 & -6.79 & -6.78 & -7.00 & -6.60 & -6.91 & -6.71 \\
    $\theta_{\mathit{se}}$  &  \cellcolor{Gray!20} 0.21 & 0.21 & 0.20     & 0.15 &  0.29 & 0.14 &  0.31 \\
    \hline
    $\bar{\ell}^{100}_{CC}(\hat{\theta}; y)$  &   \cellcolor{Gray!40} -44\,580 & -44\,584 & -44\,585  & & & & \\
    \thickhline
    \end{tabular}
    \begin{flushleft} \textsuperscript{a} Results are given for three runs of the numerical optimization algorithm. Highest estimated value for the objective function, $\bar{\ell}^{100}_{CC}(\hat{\theta}; y)$, was for run 1 (highlighted in gray). \textsuperscript{b} and \textsuperscript{c} gives endpoints of confidence intervals determined either on the basis of a shared cut-off for all the parameters or of parameter specific cut-offs. Underlined and overlined values indicates boundary of the parameter space. Definitions of the parameters are given in Table~\ref{tab:paramdef}.
    \end{flushleft}
    \label{tab:valparamestim}
    \end{adjustwidth} 
    \end{table}


\section*{Discussion}

In this work, we proposed a likelihood free inference procedure to estimate disease spread parameters at a metapopulation level. We defined two new and highly efficient objective functions based on approximate composite likelihoods which identified the correct scenario in more than 95\% of the synthetic datasets. The criteria designed correctly deal with the imperfect detection of the infection due to the lack of sensitivity of the diagnostic test and the high proportion of test-negative animals (low prevalence level in the herds).
Applying our method to serological observations from a longitudinal dataset for bovine paratuberculosis in Brittany (western France), we draw a detailed picture of infection herd statuses and of the regional spread of Map.

\textcolor{black}{TBC}

\section*{Conclusion}

The likelihood-free inference procedure presented in this paper is able to deal with limited data, such as a number of test-positive animals at a limited number of screening dates in a limited number of herds. Even with such a partial information, our method succeeds in estimating many parameters of a spatiotemporal and stochastic epidemiological models with a high accuracy. Data initially available at the individual level become finally more relevant when aggregated at herd scale.

\textcolor{black}{TBC}

\section*{Supporting information}

\paragraph*{S1 Table.}
\label{S1Table}
{\bf Performances of 16 candidate objective functions based on an LHS of 2,000 scenarios}

\paragraph*{S2 Table.}
\label{S2Table}
{\bf Numerical identifiability analysis on 20 simulated scenarios (40 simulated datasets).}

\paragraph*{S1 Fig.}
\label{S1Fig}
{\bf Comparison the approximate composite log-likelihood considering or neglecting the first-order Markovian dependence.}

\paragraph*{S2 Fig.}
\label{S2Fig}
{\bf Comparison of true and estimated initial within-herd prevalence distribution for 20 simulated scenarios (40 simulated datasets).}

\paragraph*{S3 Fig.}
\label{S3Fig}
{\bf Assessment of the impact of sub-optimal numerical optimization on the calibration of thresholds to build confidence intervals.}

\paragraph*{S4 Fig.}
\label{S4Fig}
{\bf Evolution of the parameter values across iterations of the procedure on real data (runs 1-4).}

\paragraph*{S5 Fig.}
\label{S5Fig}
{\bf Evolution of the parameter values through profiles of approximate composite likelihood.}

\section*{Acknowledgments}
This work was carried out with the financial support of the French Research Agency (ANR), Program Investments for the Future, project ANR-10-BINF-07 (MIHMES), project ANR-16-CE32-0007-01 (CADENCE), the INRAE Métaprogramme GISA, project PREDICATT, and the European Union through the European fund for the regional development (FEDER) of Pays-de-la-Loire. We thank the DGAl and GDS Bretagne for providing the data. We are grateful to the INRAE MIGALE bioinformatics facility (MIGALE, INRAE, 2020. Migale bioinformatics Facility, doi: 10.15454/1.5572390655343293E12) for providing computing resources.

\nolinenumbers

\bibliography{list_of_references}


\clearpage

\begin{sidewaystable}[!hp]
\vspace{-2.25in}
\centering
\begin{flushleft}
    {\bf S1 Table.} Performances of 16 candidate objective functions based on an LHS of 2,000 scenarios\medskip
\end{flushleft}
\setlength{\tabcolsep}{0.15em} 
\renewcommand{\arraystretch}{1.2}
\begin{tabular}{lp{0.055\textwidth}p{0.055\textwidth}p{0.055\textwidth}p{0.055\textwidth}p{0.055\textwidth}p{0.055\textwidth}p{0.055\textwidth}p{0.055\textwidth}p{0.055\textwidth}p{0.055\textwidth}p{0.055\textwidth}p{0.055\textwidth}p{0.055\textwidth}p{0.055\textwidth}p{0.055\textwidth}c}
\thickhline
                                                &  \multicolumn{16}{c}{Objective function\textsuperscript{b}}    \\
Perf.\textsuperscript{a}                            & $\tilde{f}^{100}_{ASAD}$  & $\tilde{f}^{100}_{ASADe}$ & $\tilde{f}^{100}_{ASADc}$ & $\tilde{f}^{100}_{ASADec}$ & $\tilde{f}^{100}_{ASSD}$  & $\tilde{f}^{100}_{ASSDe}$ & $\tilde{f}^{100}_{ASSDc}$ & $\tilde{f}^{100}_{ASSDec}$ & $\tilde{f}^{100}_{SSDA}$  & $\tilde{f}^{100}_{SSDAe}$ & $\tilde{f}^{100}_{SSDAc}$ & $\tilde{f}^{100}_{SSDAec}$ & $\tilde{\ell}^{100}_{BI}$    & $\tilde{\ell}^{100}_{BIe}$   & $\tilde{\ell}^{100}_{MC}$    & $\tilde{\ell}^{100}_{CC}$    \\ \thickhline
\multicolumn{17}{l}{Rank of $\theta^0$} \\
\hspace{0cm}first   & 0.195 & 0.270 & 0.355 & 0.395  & 0.170 & 0.400 & 0.265 & 0.470  & 0.897 & 0.904 & 0.594 & 0.598  & 0.754 & 0.758 & 0.970 & 0.954 \\
\hspace{0cm}top 5\%                                  & 0.416 & 0.484 & 0.467 & 0.526  & 0.369 & 0.452 & 0.431 & 0.483  & 1.00  & 1.00  & 0.994 & 0.994  & 1.00  & 1.00  & 1.00  & 1.00  \\
\hline
\multicolumn{17}{l}{Error on parameter ($|\theta^{*}_{\cdot}-\theta^0_{\cdot}|/|\theta^{\max}_{\cdot}-\theta^{\min}_{\cdot}|$)} \\
\hspace{0cm}$\theta_{hi}$  & 0.268 & 0.251 & 0.270 & 0.256  & 0.285 & 0.305 & 0.256 & 0.269  & 0.019 & 0.018 & 0.054 & 0.054  & 0.034 & 0.033 & 0.001 & 0.002 \\
\hspace{0cm}$\theta_{mi}$        & 0.260 & 0.263 & 0.264 & 0.264  & 0.259 & 0.272 & 0.257 & 0.267  & 0.027 & 0.025 & 0.115 & 0.117  & 0.072 & 0.071 & 0.006 & 0.011 \\
\hspace{0cm}$\theta_{si}$        & 0.328 & 0.320 & 0.325 & 0.317  & 0.352 & 0.339 & 0.347 & 0.335  & 0.027 & 0.025 & 0.121 & 0.120  & 0.069 & 0.067 & 0.006 & 0.009 \\
\hspace{0cm}$\theta_{mi^t}$  & 0.214 & 0.209 & 0.214 & 0.214  & 0.206 & 0.206 & 0.209 & 0.212  & 0.017 & 0.015 & 0.073 & 0.074  & 0.043 & 0.042 & 0.004 & 0.006 \\
\hspace{0cm}$\theta_{si^t}$  & 0.264 & 0.254 & 0.260 & 0.254  & 0.278 & 0.263 & 0.279 & 0.270  & 0.019 & 0.018 & 0.088 & 0.088  & 0.048 & 0.048 & 0.004 & 0.007 \\
\hspace{0cm}$\theta_{oi}$        & 0.253 & 0.270 & 0.246 & 0.253  & 0.278 & 0.288 & 0.278 & 0.287  & 0.015 & 0.015 & 0.068 & 0.067  & 0.040 & 0.038 & 0.004 & 0.005 \\
\hspace{0cm}$\theta_{od}$        & 0.240 & 0.244 & 0.219 & 0.220  & 0.254 & 0.267 & 0.238 & 0.248  & 0.016 & 0.015 & 0.064 & 0.062  & 0.039 & 0.038 & 0.005 & 0.007 \\
\hspace{0cm}$\theta_{bg}$        & 0.135 & 0.140 & 0.129 & 0.131  & 0.146 & 0.165 & 0.142 & 0.157  & 0.008 & 0.007 & 0.040 & 0.041  & 0.023 & 0.022 & 0.002 & 0.003 \\
\hspace{0cm}$\theta_{se}$        & 0.157 & 0.168 & 0.141 & 0.148  & 0.196 & 0.214 & 0.175 & 0.191  & 0.012 & 0.011 & 0.028 & 0.028  & 0.022 & 0.023 & 0.002 & 0.003 \\
\hline
\multicolumn{17}{l}{Error on parameter, after excluding the true scenario ($|\theta^{+}_{\cdot}-\theta^0_{\cdot}|/|\theta^{\max}_{\cdot}-\theta^{\min}_{\cdot}|$)} \\
\hspace{0cm}$\theta_{hi}$        & 0.269 & 0.253 & 0.272 & 0.261  & 0.287 & 0.311 & 0.261 & 0.276  & 0.157 & 0.157 & 0.155 & 0.154  & 0.152 & 0.152 & 0.080 & 0.085 \\
\hspace{0cm}$\theta_{mi}$        & 0.263 & 0.267 & 0.271 & 0.272  & 0.262 & 0.281 & 0.261 & 0.277  & 0.276 & 0.278 & 0.280 & 0.280  & 0.282 & 0.283 & 0.246 & 0.245 \\
\hspace{0cm}$\theta_{si}$        & 0.334 & 0.328 & 0.335 & 0.328  & 0.358 & 0.351 & 0.355 & 0.348  & 0.288 & 0.287 & 0.296 & 0.298  & 0.299 & 0.300  & 0.257 & 0.253 \\
\hspace{0cm}$\theta_{mi^t}$       & 0.217 & 0.214 & 0.221 & 0.222  & 0.209 & 0.215 & 0.214 & 0.222  & 0.171 & 0.172 & 0.186 & 0.186  & 0.173 & 0.175 & 0.143 & 0.139 \\
\hspace{0cm}$\theta_{si^t}$       & 0.269 & 0.261 & 0.268 & 0.264  & 0.283 & 0.273 & 0.286 & 0.280  & 0.211 & 0.211 & 0.224 & 0.225  & 0.216 & 0.217 & 0.174 & 0.170  \\
\hspace{0cm}$\theta_{oi}$        & 0.257 & 0.276 & 0.254 & 0.263  & 0.281 & 0.299 & 0.283 & 0.298  & 0.166 & 0.167 & 0.167 & 0.168  & 0.173 & 0.176 & 0.148 & 0.148 \\
\hspace{0cm}$\theta_{od}$        & 0.245 & 0.251 & 0.227 & 0.231  & 0.258 & 0.278 & 0.246 & 0.262  & 0.177 & 0.179 & 0.165 & 0.165  & 0.178 & 0.180 & 0.175 & 0.171 \\
\hspace{0cm}$\theta_{bg}$        & 0.137 & 0.144 & 0.133 & 0.136  & 0.148 & 0.170 & 0.145 & 0.164  & 0.076 & 0.077 & 0.086 & 0.087  & 0.084 & 0.085 & 0.081 & 0.077 \\
\hspace{0cm}$\theta_{se}$        & 0.159 & 0.170 & 0.144 & 0.151  & 0.197 & 0.216 & 0.178 & 0.194  & 0.083 & 0.083 & 0.077 & 0.078  & 0.082 & 0.083 & 0.059 & 0.057 \\
\thickhline
\end{tabular}
\begin{flushleft}
 \textsuperscript{a} See Table~\ref{tab:comparisoncriteriaperf} for a definition of the different performance measures. \textsuperscript{b} Objective functions: $ASAD$, average sum of absolute differences between the proportion of positive tests; $ASSD$, the average sum of squared differences between the proportion of positive tests; $SSDA$, sum of squared of differences between the average proportion of positive tests; $BI$, log-likelihood under a binomial model with independent sampling events; $MC$, marginal composite likelihood; $CC$, conditional composite likelihood. As described in the Methods section, the subscripts $e$ and $c$ denote variants of these objective functions: $e$, proportion of positive tests in each simulated trajectory is replaced by its expected value; $c$, reweighing each term of the objective function by the square of the number of tests. Values for $ASSD$, $SSDA$, $BI$, $MC$, and $CC$ are the same as those reported in Table~\ref{tab:comparisoncriteriaperf}.
\end{flushleft}
\label{tab:comparisoncriteriaperfSI}
\end{sidewaystable}

\clearpage

\begin{table}[!ht]
\begin{adjustwidth}{-2.25in}{0in} 
\begin{flushleft}
    {\bf S2 Table.} Numerical identifiability analysis on 20 simulated scenarios (40 simulated datasets).\medskip
\end{flushleft}
\setlength{\tabcolsep}{0.35em} 
\renewcommand{\arraystretch}{1.15}
\begin{tabular}{lccccccccccccccrrc}
\thickhline
\multirow{2}{*}{\#} & \multicolumn{14}{c}{True and estimated parameter values} & \multicolumn{2}{c}{Objective function} & t-test\textsuperscript{a} \\
 & $\theta_{\mathit{hi}}^\circ$ & $\theta_{\mathit{hi}}^\star$ & $\theta_{\mathit{{mi}^t}}^\circ$ & $\theta_{\mathit{{mi}^t}}^\star$ & $\theta_{\mathit{{si}^t}}^\circ$ & $\theta_{\mathit{{si}^t}}^\star$ & $\theta_{\mathit{oi}}^\circ$ & $\theta_{\mathit{oi}}^\star$ & $\theta_{\mathit{od}}^\circ$ & $\theta_{\mathit{od}}^\star$ & $\theta_{\mathit{bg}}^\circ$ & $\theta_{\mathit{bg}}^\star$ & $\theta_{\mathit{se}}^\circ$ & $\theta_{\mathit{se}}^\star$ & $\bar{\ell}^{100}_{CC}(\theta^{0}; y)$ & $\bar{\ell}^{100}_{CC}(\hat{\theta}; y)$ & p-value \\
\thickhline
1a  & 0.87 & 0.87 & 0.22 & 0.29 & 0.18 & 0.22 & 0.33 & 0.38 &  0.00 & -0.00 & -7.10 & -7.19 & 0.94 & 0.80 &  -83\,646.0 &  -83\,593.9 & -- \\
1b  & 0.87 & 0.85 & 0.22 & 0.33 & 0.18 & 0.23 & 0.33 & 0.39 &  0.00 & -0.02 & -7.10 & -7.27 & 0.94 & 0.76 &  -83\,619.2 &  -83\,572.7 & -- \\ 
2a  & 0.11 & 0.14 & 0.15 & 0.24 & 0.13 & 0.19 & 0.46 & 0.48 & -0.17 & -0.07 & -6.21 & -6.25 & 0.55 & 0.48 &  -47\,044.6 &  -46\,996.6 & -- \\
2b  & 0.11 & 0.14 & 0.15 & 0.31 & 0.13 & 0.22 & 0.46 & 0.48 & -0.17 & -0.14 & -6.21 & -6.29 & 0.55 & 0.49 &  -46\,155.0 &  -46\,173.7 & 0.078 \\ 
3a  & 0.49 & 0.61 & 0.10 & 0.09 & 0.08 & 0.08 & 0.74 & 0.60 &  0.20 &  0.30 & -4.56 & -5.23 & 0.05 & 0.06 &  -28\,140.9 &  -28\,144.7 & 0.065 \\
\rowcolor{Gray!40}
3b  & 0.49 & 0.48 & 0.10 & 0.15 & 0.08 & 0.13 & 0.74 & 0.87 &  0.20 & -0.12 & -4.56 & -5.25 & 0.05 & 0.06 &  -28\,284.0 &  -28\,291.1 & 3.6e-03\textsuperscript{b} \\ 
4a  & 0.91 & 0.84 & 0.05 & 0.07 & 0.04 & 0.07 & 0.39 & 0.59 &  0.31 &  0.12 & -6.84 & -6.83 & 0.35 & 0.28 &  -42\,249.9 &  -42\,239.5 & -- \\
4b  & 0.91 & 0.71 & 0.05 & 0.10 & 0.04 & 0.08 & 0.39 & 0.56 &  0.31 &  0.35 & -6.84 & -6.88 & 0.35 & 0.25 &  -41\,380.4 &  -41\,376.5 & -- \\ 
5a  & 0.58 & 0.61 & 0.08 & 0.12 & 0.06 & 0.10 & 0.15 & 0.17 & -0.38 & -0.45 & -6.50 & -6.50 & 0.46 & 0.33 &  -35\,280.7 &  -35\,273.7 & -- \\
5b  & 0.58 & 0.60 & 0.08 & 0.13 & 0.06 & 0.11 & 0.15 & 0.18 & -0.38 & -0.39 & -6.50 & -6.58 & 0.46 & 0.34 &  -36\,203.3 &  -36\,192.0 & -- \\ 
6a  & 0.18 & 0.23 & 0.13 & 0.13 & 0.08 & 0.12 & 0.65 & 0.84 &  0.47 &  0.22 & -5.90 & -5.95 & 0.74 & 0.67 &  -81\,869.0 &  -81\,838.5 & -- \\
6b  & 0.18 & 0.21 & 0.13 & 0.21 & 0.08 & 0.17 & 0.65 & 0.82 &  0.47 &  0.22 & -5.90 & -5.98 & 0.74 & 0.70 &  -81\,787.5 &  -81\,791.4 & 0.84 \\ 
\rowcolor{Gray!40}
7a  & 0.69 & 0.72 & 0.20 & 0.20 & 0.05 & 0.15 & 0.56 & 0.56 & -0.09 & -0.05 & -5.47 & -5.18 & 0.68 & 0.63 & -136\,112.1 & -136\,224.4 & 5.6e-12 \\
\rowcolor{Gray!40}
7b  & 0.69 & 0.76 & 0.20 & 0.20 & 0.05 & 0.14 & 0.56 & 0.58 & -0.09 & -0.02 & -5.47 & -5.24 & 0.68 & 0.63 & -136\,887.9 & -137\,010.4 & 2.2e-11 \\ 
\rowcolor{Gray!40}
8a  & 0.37 & 0.41 & 0.30 & 0.40 & 0.10 & 0.25 & 0.05 & 0.06 & -0.21 & -0.44 & -4.84 & -4.86 & 0.28 & 0.28 &  -59\,054.5 &  -59\,102.5 & 8.4e-10 \\
\rowcolor{Gray!40}
8b  & 0.37 & 0.41 & 0.30 & 0.35 & 0.10 & 0.23 & 0.05 & 0.03 & -0.21 & -0.16 & -4.84 & -5.00 & 0.28 & 0.28 &  -60\,234.7 &  -60\,267.9 & 2.1e-04 \\ 
\rowcolor{Gray!40}
9a  & 0.29 & 0.38 & 0.40 & 0.34 & 0.15 & 0.23 & 0.89 & 0.90 &  0.29 &  0.40 & -7.35 & -7.36 & 0.83 & 0.82 &  -73\,143.6 &  -73\,196.3 & 9.5e-07 \\
\rowcolor{Gray!40}
9b  & 0.29 & 0.37 & 0.40 & 0.37 & 0.15 & 0.24 & 0.89 & 0.94 &  0.29 &  0.37 & -7.35 & -7.39 & 0.83 & 0.80 &  -73\,216.9 &  -73\,251.7 & 1.7e-04 \\ 
10a & 0.71 & 0.74 & 0.31 & 0.37 & 0.21 & 0.24 & 0.19 & 0.16 & -0.43 & -0.26 & -5.11 & -5.16 & 0.22 & 0.21 &  -81\,253.8 &  -81\,254.1 & 0.97 \\
10b & 0.71 & 0.75 & 0.31 & 0.26 & 0.21 & 0.19 & 0.19 & 0.13 & -0.43 & -0.24 & -5.11 & -5.13 & 0.22 & 0.22 &  -81\,425.1 &  -81\,425.9 & 0.80 \\ 
11a & 0.88 & 0.92 & 0.05 & 0.05 & 0.05 & 0.05 & 0.03 & 0.02 & -0.45 & -0.19 & -5.83 & -5.83 & 0.39 & 0.38 &  -62\,359.3 &  -62\,342.8 & -- \\
11b & 0.88 & 0.81 & 0.05 & 0.07 & 0.05 & 0.06 & 0.03 & 0.00 & -0.45 &  0.03 & -5.83 & -5.81 & 0.39 & 0.35 &  -60\,199.2 &  -60\,184.7 & -- \\ 
12a & 0.14 & 0.15 & 0.21 & 0.35 & 0.17 & 0.23 & 0.16 & 0.18 &  0.05 &  0.04 & -6.11 & -6.08 & 0.26 & 0.21 &  -25\,199.1 &  -25\,198.4 & -- \\
12b & 0.14 & 0.15 & 0.21 & 0.35 & 0.17 & 0.23 & 0.16 & 0.18 &  0.05 &  0.04 & -6.11 & -6.18 & 0.26 & 0.23 &  -26\,313.7 &  -26\,292.4 & -- \\ 
13a & 0.59 & 0.62 & 0.33 & 0.33 & 0.22 & 0.23 & 0.82 & 0.88 &  0.32 &  0.15 & -4.98 & -5.13 & 0.62 & 0.62 & -129\,600.9 & -129\,587.4 & -- \\
13b & 0.59 & 0.59 & 0.33 & 0.35 & 0.22 & 0.24 & 0.82 & 0.88 &  0.32 &  0.28 & -4.98 & -5.01 & 0.62 & 0.61 & -129\,417.4 & -129\,403.8 & -- \\ 
\rowcolor{Gray!40}
14a & 0.30 & 0.31 & 0.38 & 0.37 & 0.04 & 0.20 & 0.70 & 0.72 & -0.25 & -0.25 & -4.54 & -4.30 & 0.15 & 0.15 &  -54\,441.6 &  -54\,456.5 & 4.9e-04 \\
\rowcolor{Gray!40}
14b & 0.30 & 0.32 & 0.38 & 0.37 & 0.04 & 0.24 & 0.70 & 0.69 & -0.25 & -0.12 & -4.54 & -4.57 & 0.15 & 0.15 &  -53\,050.9 &  -53\,065.8 & 1.9e-04 \\ 
15a & 0.61 & 0.63 & 0.37 & 0.37 & 0.24 & 0.24 & 0.20 & 0.25 &  0.14 &  0.09 & -5.52 & -5.55 & 0.50 & 0.49 & -111\,653.4 & -111\,649.2 & -- \\
15b & 0.61 & 0.61 & 0.37 & 0.36 & 0.24 & 0.24 & 0.20 & 0.24 &  0.14 &  0.06 & -5.52 & -5.44 & 0.50 & 0.49 & -110\,144.8 & -110\,143.5 & -- \\ 
\rowcolor{Gray!20}
16a & 0.33 & 0.37 & 0.04 & 0.04 & 0.04 & 0.04 & 0.76 & 0.90 &  0.43 & -0.04 & -6.31 & -6.39 & 0.07 & 0.07 &  -17\,779.5 &  -17\,783.8 & 0.047 \\
\rowcolor{Gray!20}
16b & 0.33 & 0.24 & 0.04 & 0.07 & 0.04 & 0.06 & 0.76 & 0.93 &  0.43 &  0.11 & -6.31 & -6.46 & 0.07 & 0.07 &  -17\,559.8 &  -17\,565.2 & 0.015 \\ 
17a & 0.76 & 0.77 & 0.15 & 0.13 & 0.12 & 0.11 & 0.30 & 0.37 &  0.26 &  0.17 & -5.11 & -5.03 & 0.83 & 0.82 & -130\,882.6 & -130\,876.9 & -- \\
17b & 0.76 & 0.81 & 0.15 & 0.15 & 0.12 & 0.12 & 0.30 & 0.46 &  0.26 &  0.01 & -5.11 & -5.29 & 0.83 & 0.82 & -132\,446.7 & -132\,430.8 & -- \\ 
18a & 0.96 & 0.84 & 0.03 & 0.05 & 0.03 & 0.04 & 0.53 & 0.67 & -0.38 & -0.37 & -7.45 & -7.44 & 0.43 & 0.30 &  -24\,823.9 &  -24\,793.3 & -- \\
18b & 0.96 & 0.67 & 0.03 & 0.08 & 0.03 & 0.07 & 0.53 & 0.85 & -0.38 & -0.55 & -7.45 & -7.66 & 0.43 & 0.24 &  -23\,486.6 &  -23\,472.8 & -- \\ 
19a & 0.45 & 0.46 & 0.51 & 0.51 & 0.21 & 0.24 & 0.39 & 0.43 & -0.02 &  0.04 & -6.78 & -6.76 & 0.86 & 0.78 &  -88\,332.4 &  -88\,304.6 & -- \\
19b & 0.45 & 0.50 & 0.51 & 0.44 & 0.21 & 0.25 & 0.39 & 0.47 & -0.02 & -0.10 & -6.78 & -6.75 & 0.86 & 0.80 &  -91\,585.6 &  -91\,583.8 & -- \\ 
20a & 0.01 & 0.03 & 0.37 & 0.33 & 0.22 & 0.23 & 0.61 & 0.78 & -0.17 & -0.21 & -7.10 & -7.23 & 0.73 & 0.58 &  -36\,395.3 &  -36\,347.8 & -- \\
20b & 0.01 & 0.04 & 0.37 & 0.27 & 0.22 & 0.21 & 0.61 & 0.72 & -0.17 & -0.05 & -7.10 & -7.10 & 0.73 & 0.54 &  -36\,581.6 &  -36\,392.5 & -- \\
\thickhline
\end{tabular}
\begin{flushleft}
  \textsuperscript{a} t-test comparing the 10 evaluations of $\tilde{\ell}^{100}_{CC}(\theta;y)$ used to compute $\bar{\ell}^{100}_{CC}(\theta;y)$ for each $\theta$, when $\bar{\ell}^{100}_{CC}(\hat{\theta};y) < \bar{\ell}^{100}_{CC}(\theta^{0};y)$.\\
  \textsuperscript{b} Particular case for scenario 3b where $\bar{\ell}^{100}_{CC}(\theta^{0}; y)  >  \bar{\ell}^{100}_{CC}(\hat{\theta}; y)$, but $\max(\tilde{\ell}^{100}_{CC}(\theta^{0}; y)) < \max(\tilde{\ell}^{100}_{CC}(\hat{\theta}; y))$, considering all evaluation of the objective function for $\hat{\theta}$ (the value obtain during the optimisation process and the 10 additional evaluations of $\tilde{\ell}^{100}_{CC}(\hat{\theta}; y)$).
\end{flushleft}
\label{tab:identifiability_analysis_results}
\end{adjustwidth}
\end{table}

\begin{figure}[!p]
\begin{adjustwidth}{-2.25in}{0in}
\begin{center}
\includegraphics[angle=0,width=0.6\paperwidth]{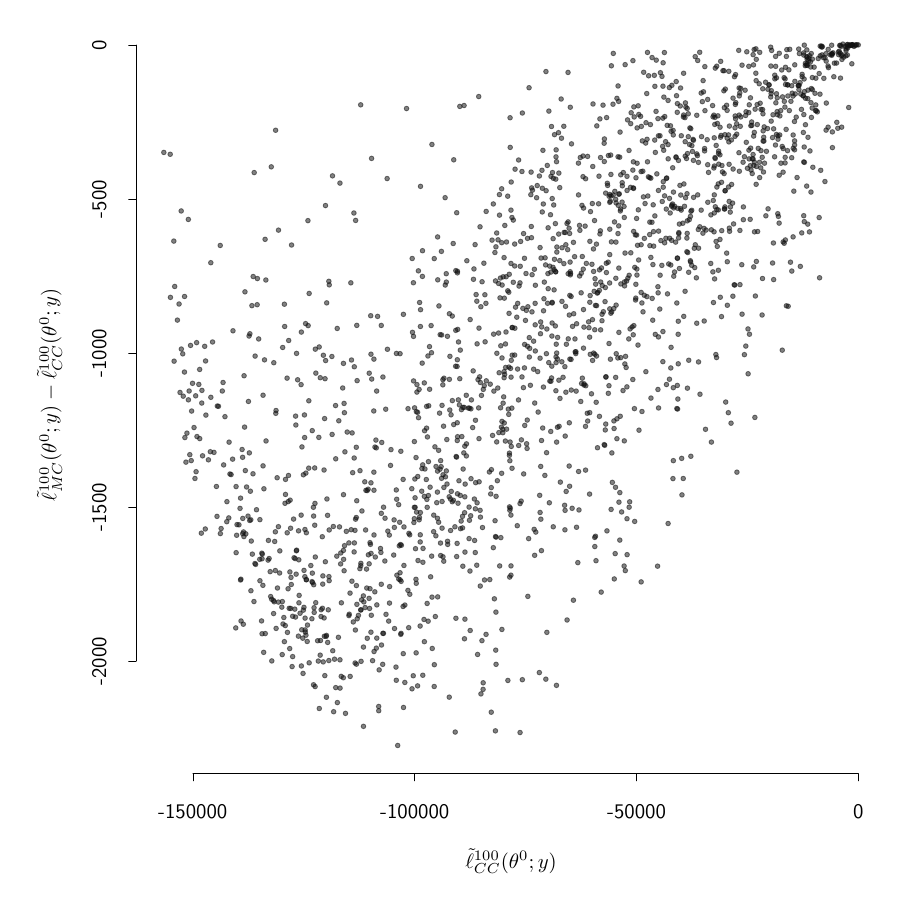}
\end{center}
{\bf S1 Fig.} Comparison the approximate composite log-likelihood
considering or neglecting the first-order Markovian
dependence. Dependence between consecutive sampling points is taken
into account in $\ell_{CC}(\theta; y)$ but not in $\ell_{MC}(\theta;
y)$.
\end{adjustwidth}
\end{figure}

\begin{sidewaysfigure}[!hp]
\vspace{-2.25in}
\includegraphics[width=0.99\textheight]{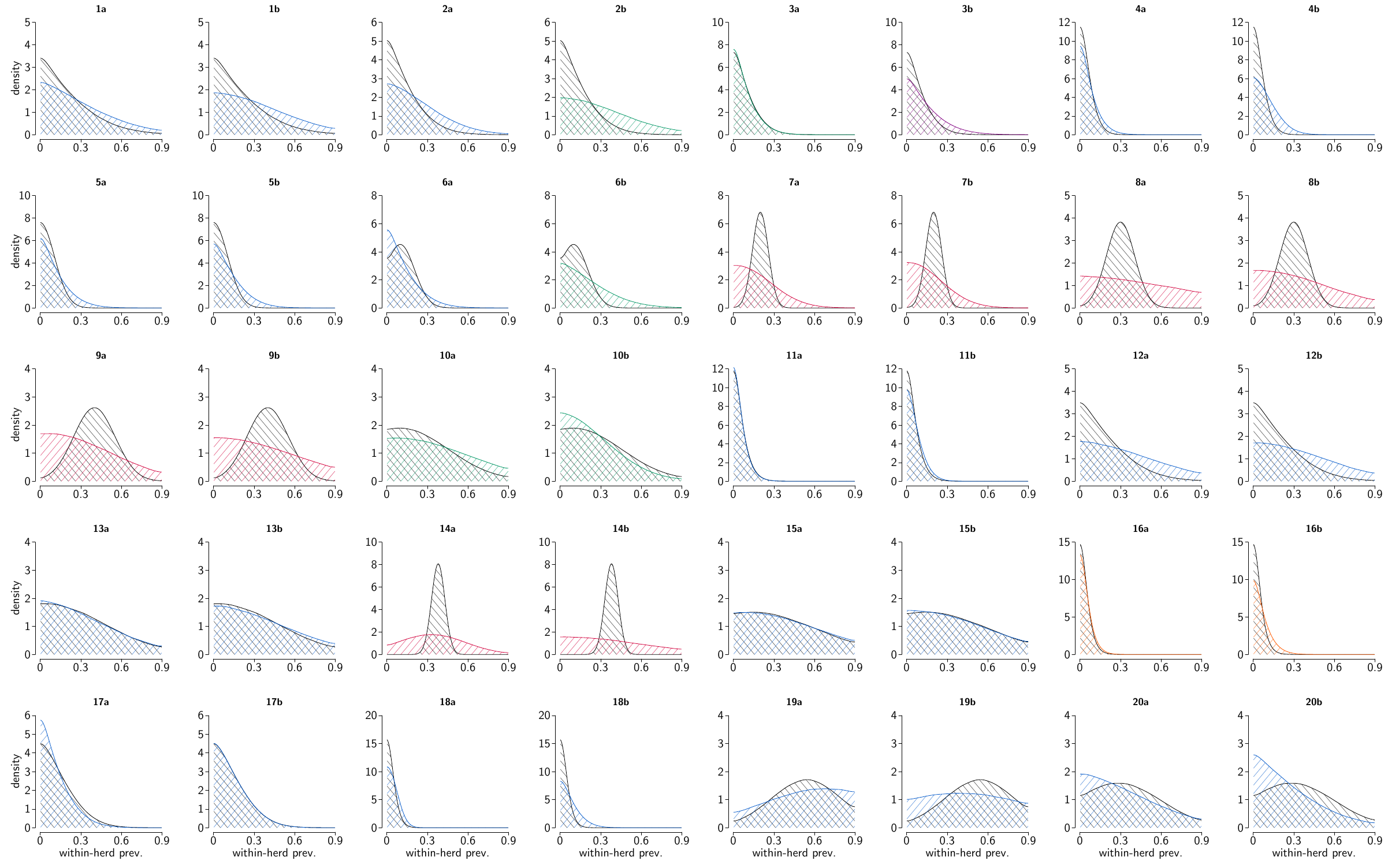}\\ \bigskip

\begin{flushleft}
  {\bf S2 Fig.} Comparison of true and estimated initial within-herd
  prevalence distribution for 20 simulated scenarios (40 simulated datasets).
  (color: blue when $\bar{\ell}^{100}_{CC}(\hat{\theta};y) >
  \bar{\ell}^{100}_{CC}(\theta^0;y)$, green or yellow when the difference is
  not statistically different, respectively, at the 5\% or 1\% significance
  level as assessed with a t-test comparing the 10 evaluations of
  $\tilde{\ell}^{100}_{CC}(\theta;y)$ used to compute
  $\bar{\ell}^{100}_{CC}(\theta;y)$ for each $\theta$, red otherwise, except for scenario 3b in purple for the reason mention in S2~Table).
  \label{sifig:comp_within_herd_prev}
\end{flushleft}
\end{sidewaysfigure}

\begin{figure}[!ht]
\begin{adjustwidth}{-2.25in}{0in}
\begin{center}
\includegraphics[width=0.65\paperwidth]{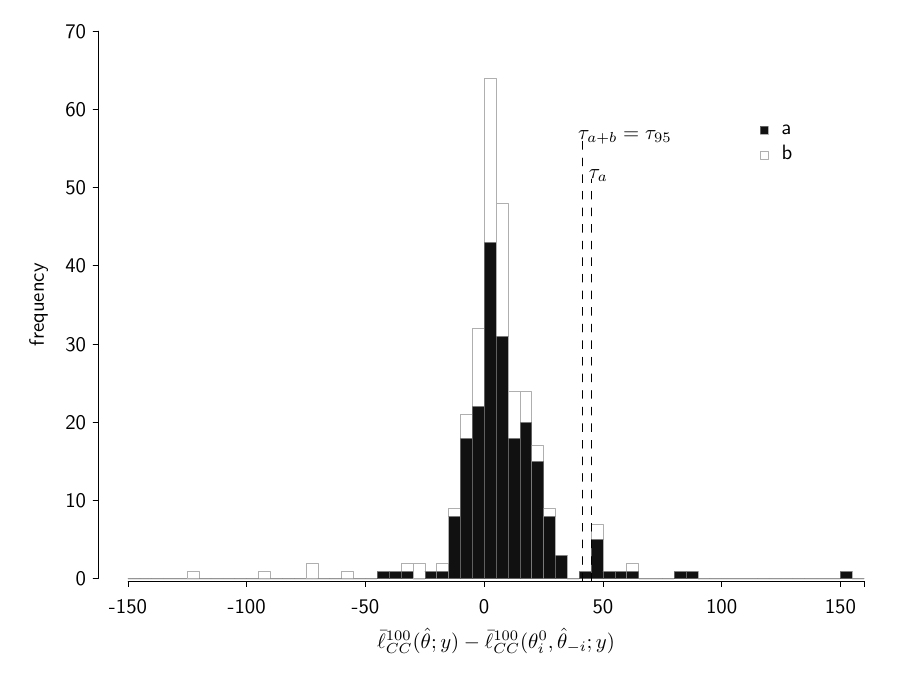}
\end{center}
{\bf S3 Fig.} Assessment of the impact of sub-optimal numerical
optimization on the calibration of thresholds to build confidence
intervals. The empirical distribution of
$\bar{\ell}^{100}_{CC}(\hat{\theta};y)-\bar{\ell}^{100}_{CC}(\theta_i^0,\hat{\theta}_{-i};y)$
that served to calibrate the threshods is represented. Black and white
areas distinguish datasets according to the sign of
$\bar{\ell}^{100}_{CC}(\hat{\theta};y)-\bar{\ell}^{100}_{CC}(\theta^0;y)$:
black when positive or when the distribution of $\tilde{\ell}^{100}_{CC}(\hat{\theta};y)$ is not statistically different from the 10 evaluations of $\tilde{\ell}^{100}_{CC}(\theta^{0};y)$, at the 5\% significance level as assessed with a t-test, white otherwise (see
Fig. \ref{fig:pointestimates_on_simulateddata}). Quantiles at level
$95\%$ of the total distribution and of the distribution excluding
white areas are reported.
\end{adjustwidth}
\end{figure}

\begin{figure}[!ht]
\begin{adjustwidth}{-2.25in}{0in}
\includegraphics[width=0.85\paperwidth]{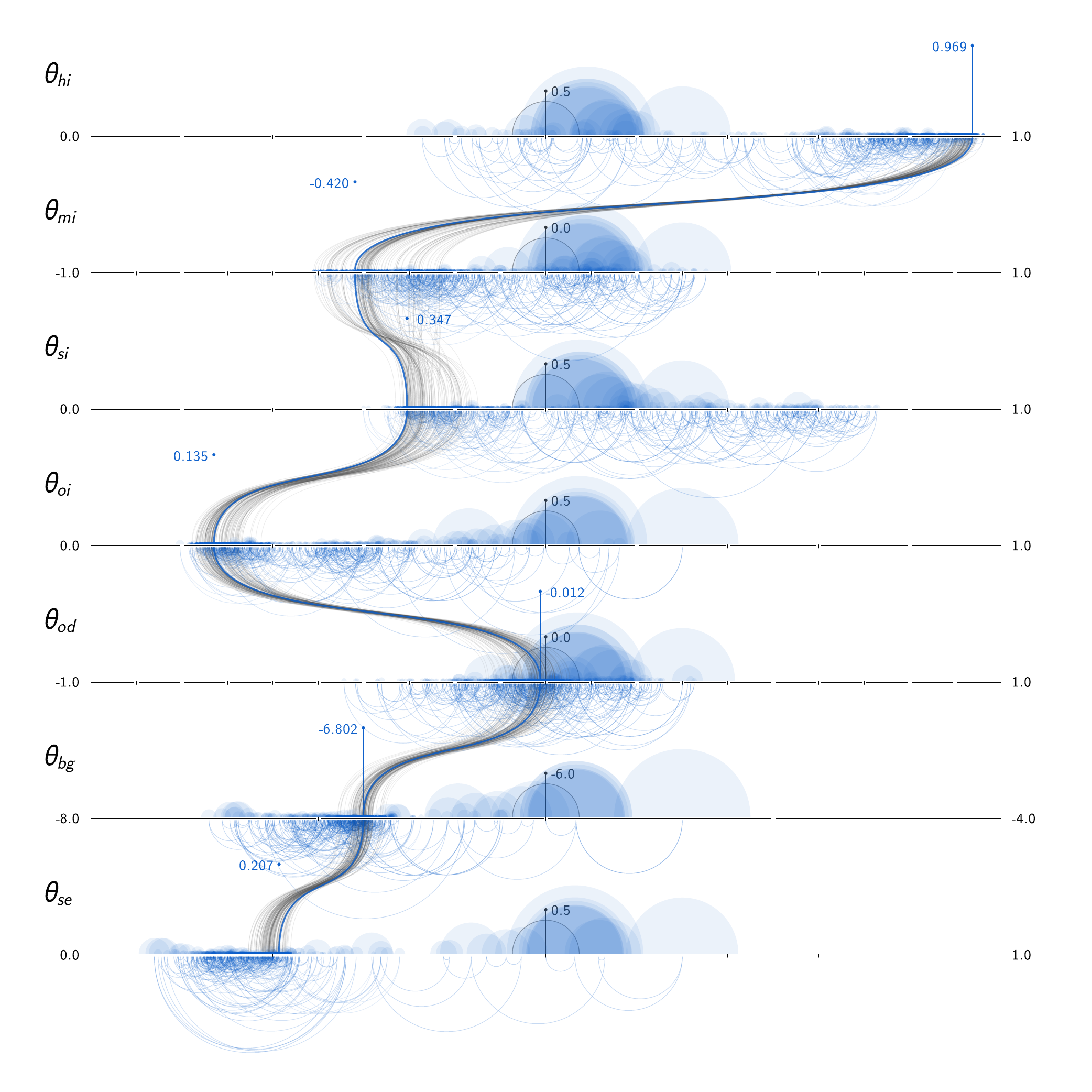}\\ {\bf
  S4 Fig. (part 1/4)} Run 1 of numerical optimization on real data. Evolution of the
parameter values across iterations of the procedure. The half colored
discs (above the x-axis) correspond to the quality of the estimation
with a size proportional to the distance from the best trajectory (the
smaller the radius, the better the value of the criterion). The half
discs with border colored in gray correspond to the initial values
(reported) used for the estimation.  The arcs (below the x-axis) link
successive evaluations. The values of the parameters associated with
the highest value of the objective function are reported, and
highlighted using a thicker colored line. The gray curved lines link
the values of the 7 parameters associated with evaluations of the
objective function above $\bar{\ell}^{100}_{CC}(\hat{\theta};y)-50$.
\label{sifig:run1}
\end{adjustwidth}
\end{figure}


\begin{figure}[!ht]
\begin{adjustwidth}{-2.25in}{0in}
  \includegraphics[width=0.85\paperwidth]{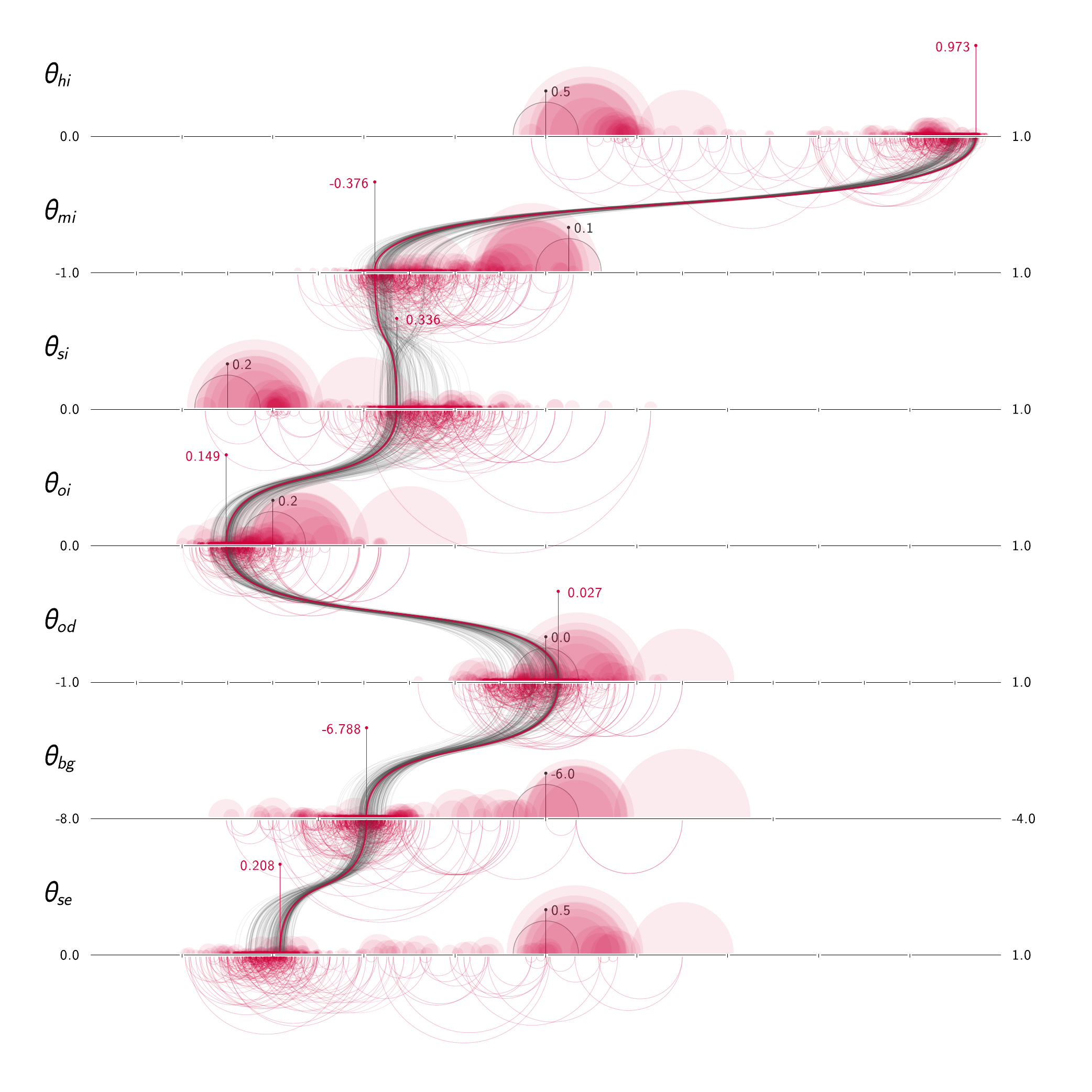} \\

  {\bf S4 Fig. (part 2/4)} Run 2 of numerical optimization on real data. Legend on page~\pageref{sifig:run1}.
\end{adjustwidth}
\end{figure}

\begin{figure}[!ht]
\begin{adjustwidth}{-2.25in}{0in}
  \includegraphics[width=0.85\paperwidth]{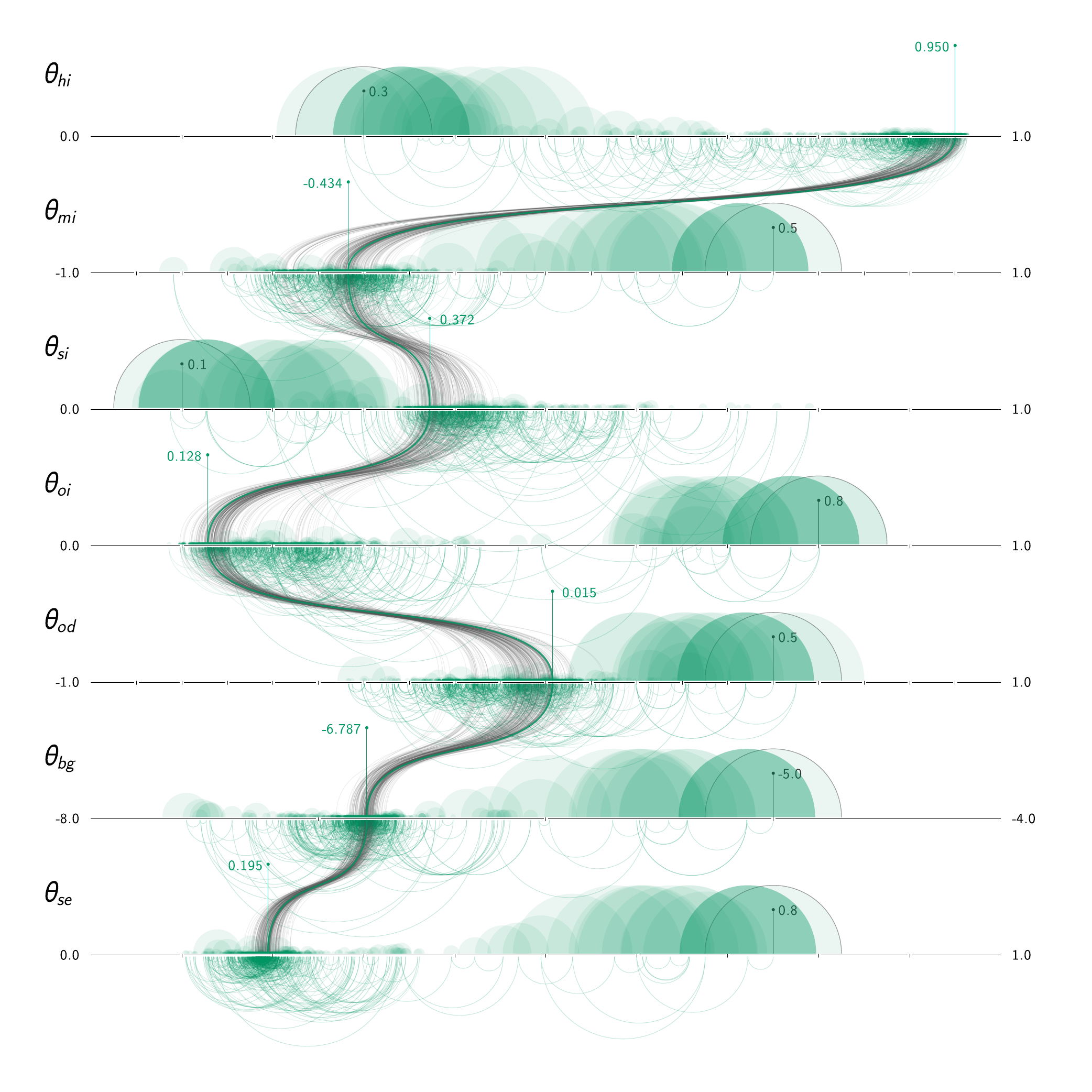}\\

  {\bf S4 Fig. (part 3/4)} Run 3 of numerical optimization on real data. Legend on page~\pageref{sifig:run1}.
  \end{adjustwidth}
\end{figure}

\begin{figure}[!ht]
\begin{adjustwidth}{-2.25in}{0in}
  \includegraphics[width=0.85\paperwidth]{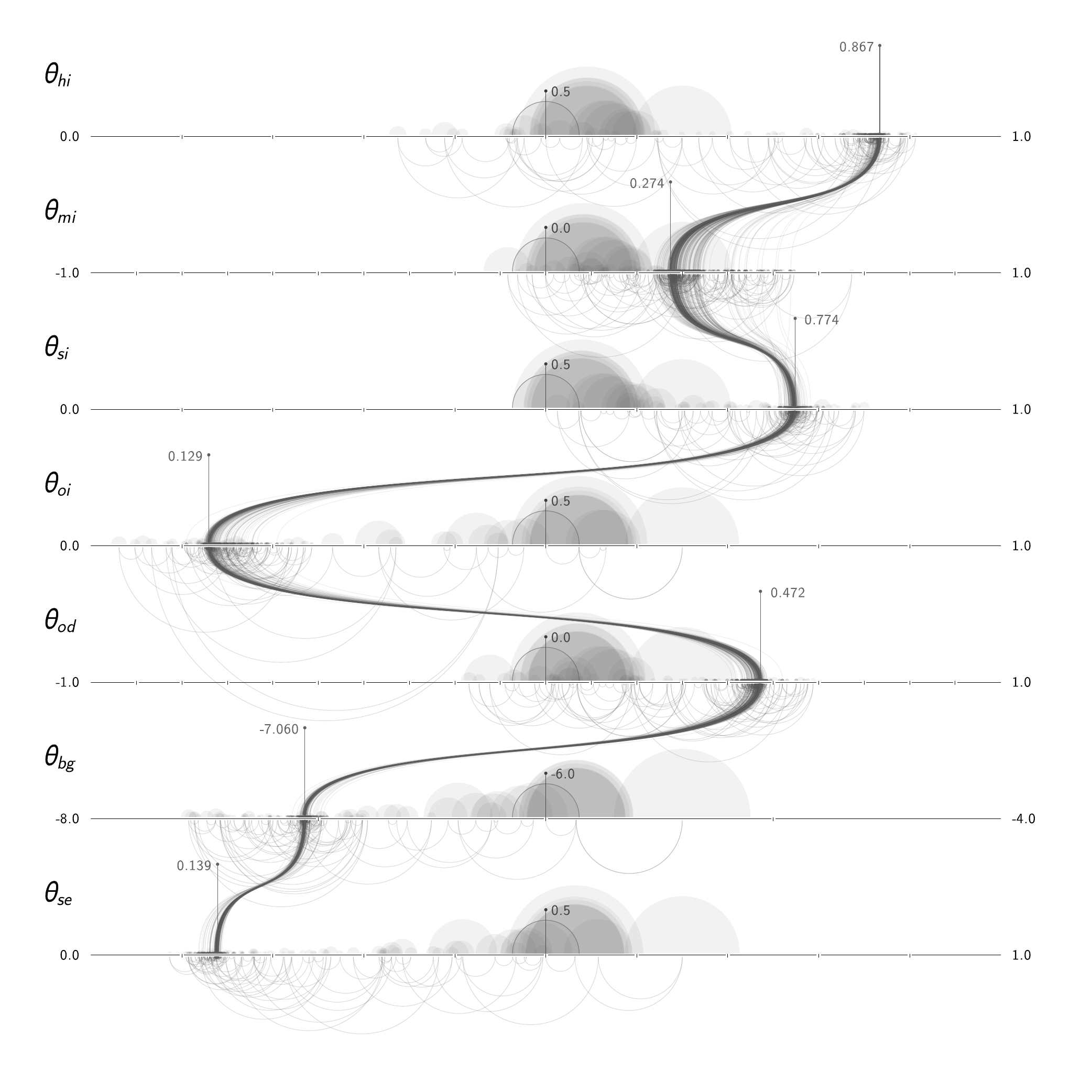}\\

{\bf S4 Fig. (part 4/4)} Run 4 of numerical optimization on real data. Legend on page~\pageref{sifig:run1}.
\end{adjustwidth}
\end{figure}


\begin{figure}[!hp]
\begin{adjustwidth}{-2.25in}{0in}
\includegraphics[angle=0,width=0.87\paperwidth]{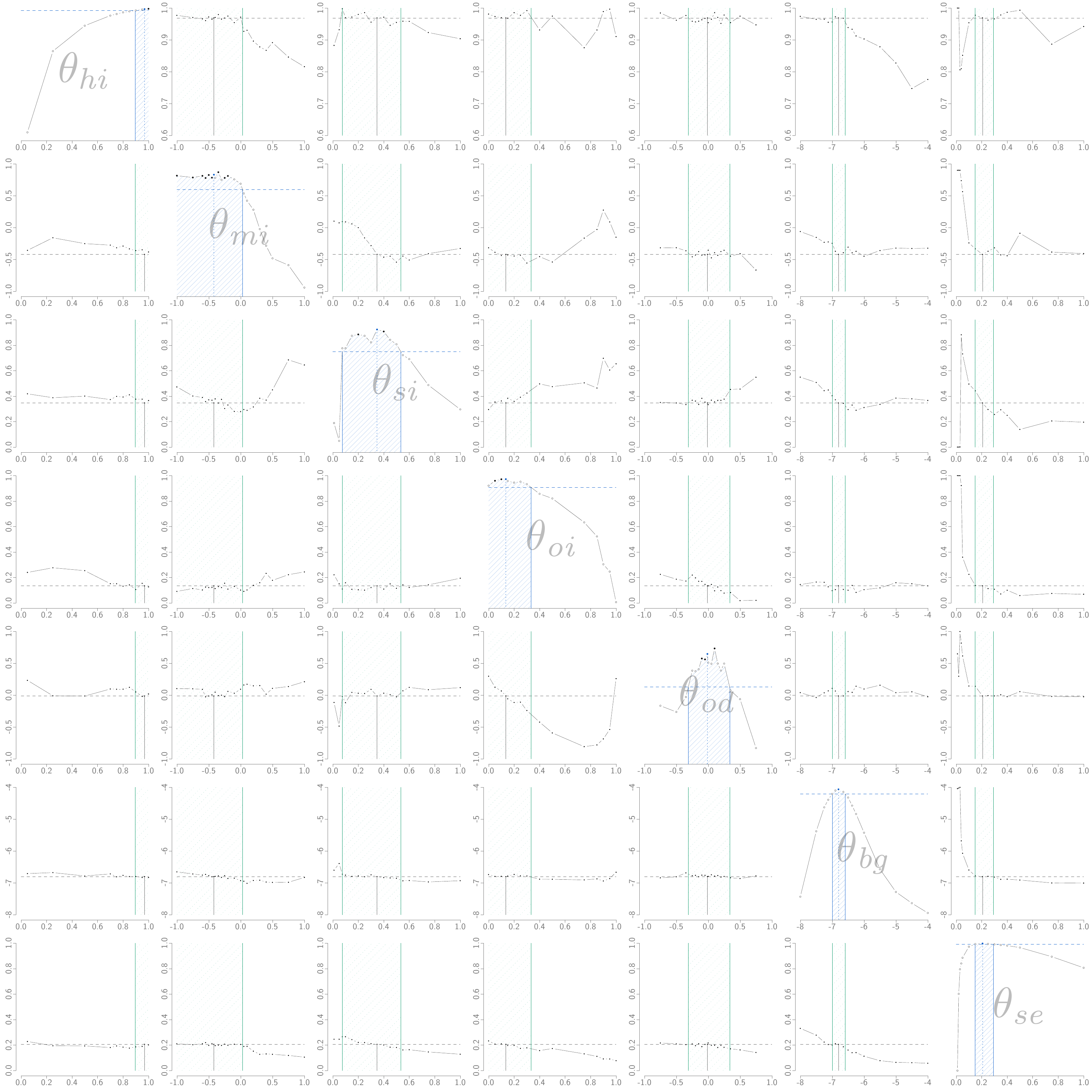}\\
                {\bf S5 Fig.} Evolution of the parameter values through profiles of approximate composite likelihood. Building the profiles used to establish confidence intervals for a parameter of interest involved numerical optimization over six other parameters at each point of a grid. The values obtained for these parameters are reported here.
\label{suppfig:profilelikelihood_param}
\end{adjustwidth}
\end{figure}


\end{document}